\documentclass[aps,prx,twocolumn,superscriptaddress,showpacs,superscriptaddress]{revtex4-2}

\usepackage[english]{babel}
\usepackage[utf8]{inputenc}
\usepackage[colorlinks=true,citecolor=blue]{hyperref}
\usepackage{graphicx,color}
\usepackage{amssymb,amsmath,amsbsy,bm,amscd}
\usepackage{epsf,epsfig}
\usepackage[shortlabels,inline]{enumitem}
\usepackage{hyperref} 
\usepackage{url}
\usepackage{physics}

\hypersetup{
    colorlinks,
    linkcolor={red},
    citecolor={red},
    urlcolor={blue}
}

\begin{document}
\title{Non-Randomness of Google's Quantum Supremacy Benchmark}
\author{Sangchul Oh}
\email{oh.sangchul@gmail.com}
\affiliation{Department of Chemistry, Department of Physics and Astronomy, and Purdue Quantum Science and
    Engineering Institute, Purdue University, West Lafayette, IN, USA}
\author{Sabre Kais}
\email{Corresponding author: kais@purdue.edu}
\affiliation{Department of Chemistry, Department of Physics and Astronomy, and Purdue Quantum Science and
    Engineering Institute, Purdue University, West Lafayette, IN, USA}
\date{\today}

\begin{abstract}
The first achievement of quantum supremacy has been claimed recently by Google for the random quantum 
circuit benchmark with 53 superconducting qubits. Here, we analyze the randomness of Google's quantum 
random-bit sampling. The heat maps of Google's random bit-strings show stripe patterns at specific qubits  
in contrast to the Haar-measure or classical random-bit strings. Google's data contains more bit 0 
than bit 1, i.e., about 2.8\% difference, and fail to pass the NIST random number tests, while the Haar-measure 
or classical random-bit samples pass. Their difference is also illustrated by the Marchenko-Pastur 
distribution and the Girko circular law of random matrices of random bit-strings.
The calculation of the Wasserstein distances shows that Google's random bit-strings are farther away from 
the Haar-measure random bit-strings than the classical random bit-strings. 
Our results imply that random matrices and the Wasserstein distance could be new tools for analyzing 
the performance of quantum computers.

\end{abstract}
\maketitle

Quantum computers could simulate nature better than classical computers, as Feynman initiated the idea of quantum 
computing~\cite{Feynman1982}. Quantum supremacy~\cite{Preskill2012,Harrow2017} that a quantum computer could perform 
certain computational tasks exponentially faster than a classical computer, is one of key milestones in developing 
practical quantum computers. The power of a quantum computer is believed to stem from its quantum nature such as 
interference, entanglement, and a large Hilbert space growing exponentially with the number of qubits. 
The speed-up of quantum algorithms such as Shor's factoring algorithm~\cite{Shor1997} or the Harrow-Hassidim-Lloyd 
algorithm for solving linear systems of equations~\cite{HHL2009} requires a large-scale and error-corrected quantum 
computer. With noisy-intermediate scale quantum computers available these days, quantum sampling algorithms are 
considered good candidates to demonstrate quantum supremacy or quantum advantage~\cite{Bouland2019}.

Recently, the achievements of quantum supremacy for quantum sampling algorithms on noisy intermediate scale 
quantum computers have been reported. In 2019, Google team~\cite{Arute2019} claimed the first quantum supremacy 
by implementing random quantum circuits on 53 superconducting qubits. More recently, Wu {\it et al.} performed
the random quantum circuits with 56 superconducting qubits~\cite{Wu2021}. 
In 2020, Zhong {\it et al.}~\cite{Zhong2020} reported the quantum advantage in the Gaussian boson 
sampling on linear optical quantum computers. The boson sampling task is to sample bit strings from 
the probability distribution of bosons, given by the permanent of a unitary 
operator~\cite{Troyansky96,Aaronson2013}.

Google's quantum supremacy benchmark task is to generate random bit-strings by applying random quantum circuits 
on qubits followed by the measurement. The probability distribution of random bit-strings generated 
by random quantum circuits is not given by a uniform random distribution, but obeys the eigenvector distribution 
of a circular unitary ensemble~\cite{Dyson1962,Kus1988}. Google's Sycamore quantum processor could generate 
millions of these random bit-strings of size $n=53$ in about 200 seconds, while a supercomputer 
with the currently known efficient classical algorithms would take much longer time~\cite{Pednault2019,Huang2020}.
To verify that a quantum computer implements random quantum circuits correctly,
the linear cross-entropy benchmark was introduced~\cite{Arute2019,Neill2018}.
The linear cross-entropy fidelity is calculated using a probability distribution obtained 
from a classical computer and output bit-strings of a quantum computer.
Its value was slightly greater than the theoretical threshold.
However, there are important questions unexplored whether Google's quantum random bit-strings are truly 
random, or how far away Google's random bit-strings are from classical random bit-strings or from 
the Haar-measure sampling. A rigorous analysis is needed to quantify the performance of quantum random 
circuits because random unitary dynamics is essential in 
chaotic scattering~\cite{Haake1990,Blumel1990,Beenakker1997,Haake2010,Livan}, 
quantum information processing~\cite{Emerson2003}, randomized benchmarking of noisy quantum 
gates~\cite{Emerson2005,Knill2008,Magesan2011}, scrambling of information in black holes and 
quantum many-body systems~\cite{Hayden2007,Nahum2018}, and hydrodynamic simulation~\cite{Richter2021} 
in addition to the quantum supremacy benchmark test.

In this paper, we analyze the randomness of output bit-strings of random quantum circuits 
using the random matrix theory~\cite{Mehta2004,Livan,Tao2015}, the NIST random number 
test code~\cite{NIST2010}, and the Wasserstein distance~\cite{Villani2008}. To compare with
the dataset of Google's quantum supremacy experiment, classical random bit-strings and the Haar-measure 
random bit-strings are generated. The heat map patterns of random bit-strings and the NIST random number 
tests will show the non-randomness of Google's random bit-strings and uncover errors of Google's random 
quantum circuits. The difference between Google' and classical random bit-strings is illustrated by 
the positions of outliers of random matrices of random bit-strings. Finally, we'll calculate
the Wasserstein distance between various data sets of random bit-strings.
It will be shown that Google's random bit strings are farther away from the Haar-measure random-bit 
strings than the classical random bit-strings 

\smallskip
\paragraph*{\bf Random Quantum Circuits\\}

Let us start with a brief introduction to the random quantum circuit benchmark and how to veryfy
its faithful implementation. The random quantum circuit benchmark starts with 
samping a random unitary operator $U(2^n)$, then applying it on an input state $\ket{0^n}$ of $n$ qubits 
and measuring the output state $\ket{\psi} = U\ket{0^n}$ in the computational basis $\{\ket{x}\}$ 
to generate the random bit string $x = a_0 a_1\cdots a_{n-1}$ with $a_i\in\{0,1\}$. 
The probability of getting a random bit string $x$ 
is given by $p_x \equiv |\braket{x}{\psi}|^2 = |\bra{x}U\ket{0^n}|^2$. 
By repeating this process $M$ times, a $M\times n$ random-bit array is obtained. 

The first key element of the random quantum circuit is how to draw unitary operators uniformly 
and randomly. Mathematically, this could be done with the Haar invariant measure on a $U(2^n)$ unitary group. 
The collection of these random unitary operators is called a circular unitary ensemble (CUE) introduced by 
Dyson~\cite{Dyson1962}. The Haar-measure sampling of a unitary operator out of the unitary group is challenging. 
A unitary operator $U(2^n)$ can be decomposed into the $(2^n-1)!$ product of 2-dimensional unitary transformations, 
called the Hurwitz decomposition~\cite{Hurwitz1963,Zyczkowski1994}. However, this decomposition requires 
a huge amount of gate operations: the number of 1 or 2-qubit gates is $n^2\times 2^{2n}$ and the number of parameters 
is $2^{2n}$. Emerson {\it et al.}~\cite{Emerson2003} proposed a method of generating pseudo-random 
unitary operators: the quantum circuit of $n$ random unitary rotations on single qubits with $3n$ parameters 
followed by the simultaneous two-body interactions on $n$ qubits are repeated $m$ times. 
On a classical computer, random unitary matrices can be generated by the QR decomposition of matrices with 
Gaussian random complex elements~\cite{Mezzadri2006}. 
The QR algorithm needs $O((2^{n})^3)$ floating point operations.

The second key element of the random quantum circuit is the statisical property of the probability $p_x$ finding 
a qubit state in $\ket{x}$. Random unitary operators make qubit states distributed uniformly in a $2^n$ 
dimensional Hilbert space. So amplitudes $c_x$ of $\ket{\psi} = \sum_x c_x\ket{x}$ may have the Gaussian 
distribution on the surface of a $2\times 2^n$ dimensional sphere. This leads to 
calculate the probability distribution $P(p)$ for the random variable $p$ (dropping the subscript $x$ of $p_x$)
\begin{align}
P(p) = (N-1)(1-p)^{N-2}\,.
\label{dist_eigenvector_CUE}
\end{align}
This is the chi-square distribution wiht 2 degrees of freedom $\chi^2_2(p)$, and 
known as the eigenvector distribution of a circular unitary ensemble~\cite{Kus1988,Haake1990,Haake2010},
For large $N$, it becomes $P(p) = Ne^{-Np}/(1-e^{-N})\approx Ne^{-Np}$. Note that this should not be 
confused with the Porter-Thomas distribution $P_{\rm PT}(q) = \frac{1}{\sqrt{2\pi Nq}} e^{-Nq/2}$
that is the eigenvector distribution of an orthogonal circular 
ensemble~\cite{Porter1956,Haake2010,Boixo2018,Neill2018,Arute2019}.
The expectation value of finding $p$ with respect to $P(p)$ is given by $1/N$. 
This is consistent with our intuition that classically the probability $p_x$ of finding a bit string $x$ 
out of $N$ possible bit strings is $1/N$. 

The last key element is to verify the faithful implementation of random quantum circuits. To this end, one has to 
estimate the empirical probability $p_x$ of finding a random quantum state in $\ket{x}$ with $x=0,\dots,2^n-1$ 
from the output data, a $M\times n$ random binary array. Then one has to construct an empirical 
probability $P_{\rm em}(p)$ of probabilities $p$ to compare it to the ideal probability $P(p)$ given by 
Eq.~(\ref{dist_eigenvector_CUE}). For a small number of qubits, the Kullback-Leibler divergence or the cross 
entropy of $P_{\rm em}(p)$ from $P(p)$ were used to measure how the probability distribution $P_{\rm em} (p)$ 
is close to the ideal distribution $P(p)$~\cite{Boixo2018,Neill2018}.

However, in Google's quantum supremacy experiment with $n=53$ qubits, the reconstruction of $P_{\rm em}$ 
was impossible because a few millions of random bit-strings is too small to estimate $p(x)$
with $x = 0 \sim 9\times 10^{15}$. Instead, the linear cross-entropy fidelity 
$F_{\rm XEB} = 2^n\cdot \frac{1}{M}\sum_{i=1}^M p(x_i) -1$ is 
introduced~\cite{Boixo2018,Bouland2019,Arute2019,Aaronson2017,Aaronson2020}.
Here, $x_i$ are the observed bit strings and the probability $p(x)$ is calculated using 
the Schr\"odinger-Feyman simulation of a single random unitary operator for $n=53$ on 
on a supercomputer. A tricky point is that Google's experiment applied the same single random 
unitary operator, rather than the CUE, to get a few millions of random bit-strings
and to calculate $p(x)$ on a supercomputer. If the CUE were used, $p(x)$ would be random
and the linear cross-entropy would not be applied.
Google obtained $F_{\rm XEB} = 0.00224$ for $n=53$ qubits.

\smallskip
\paragraph*{\bf Non-randomness of Google's Random Bit Strings\\} 

\begin{figure*}[t]
\includegraphics[width=0.32\textwidth]{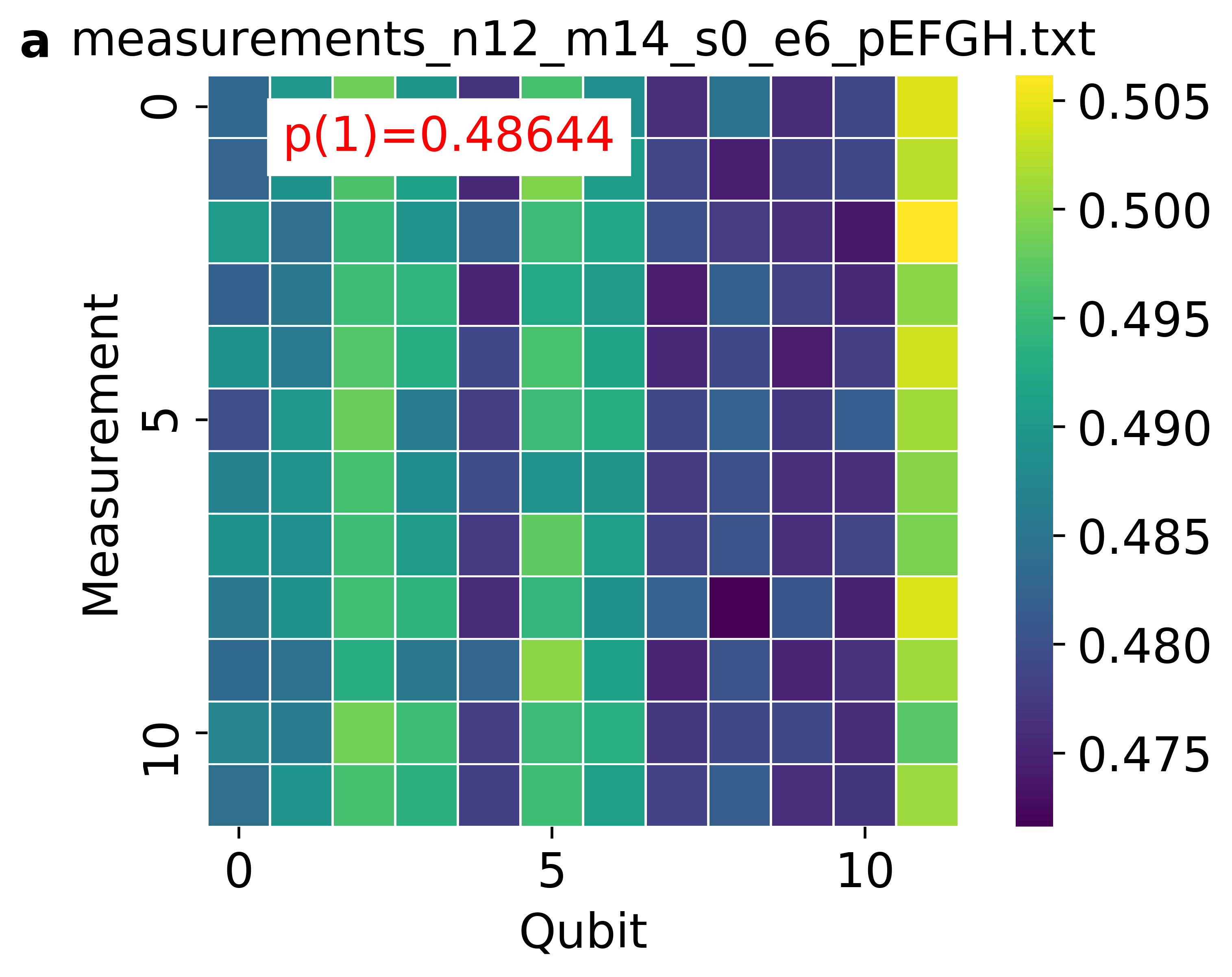}
\includegraphics[width=0.32\textwidth]{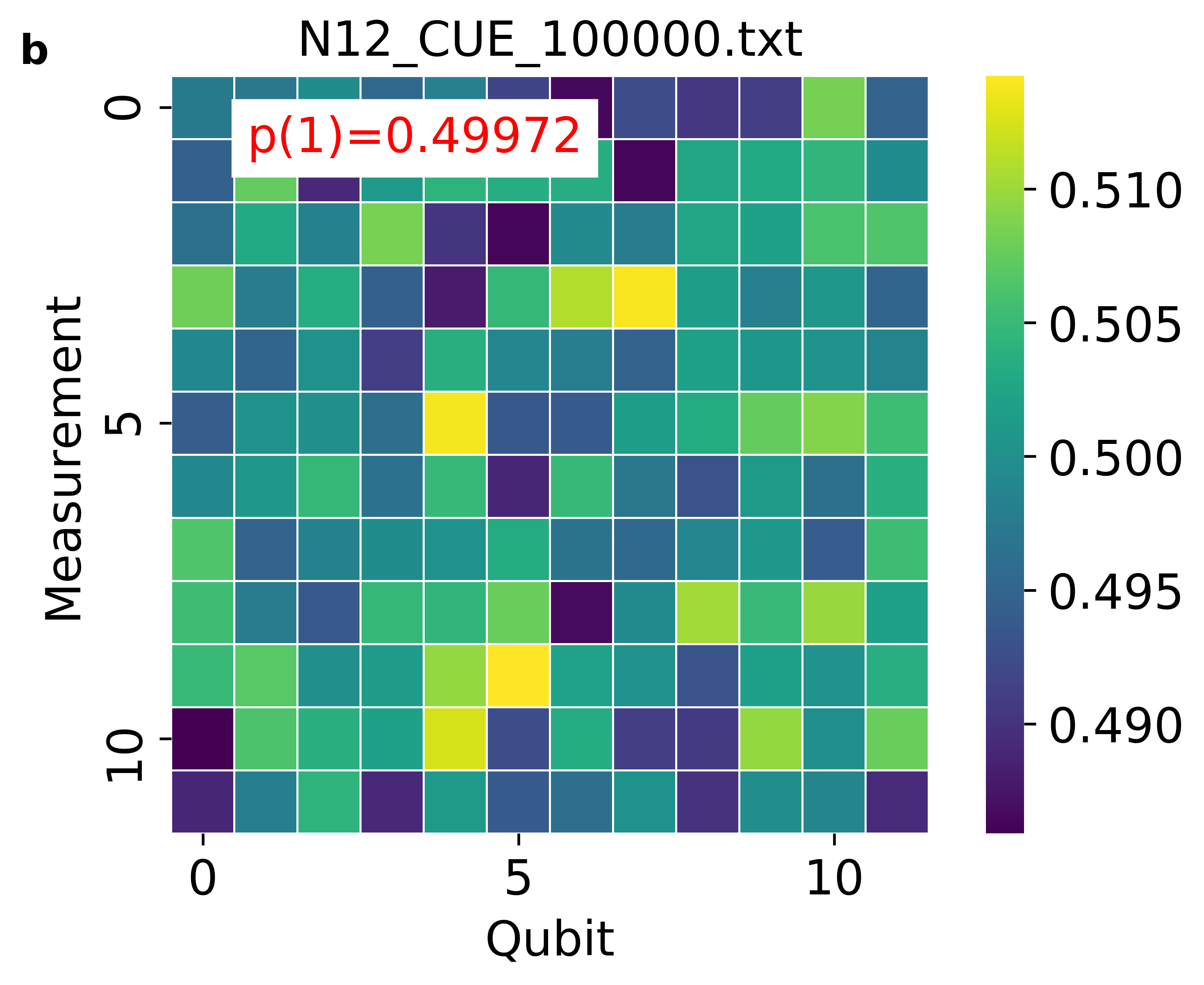}
\includegraphics[width=0.32\textwidth]{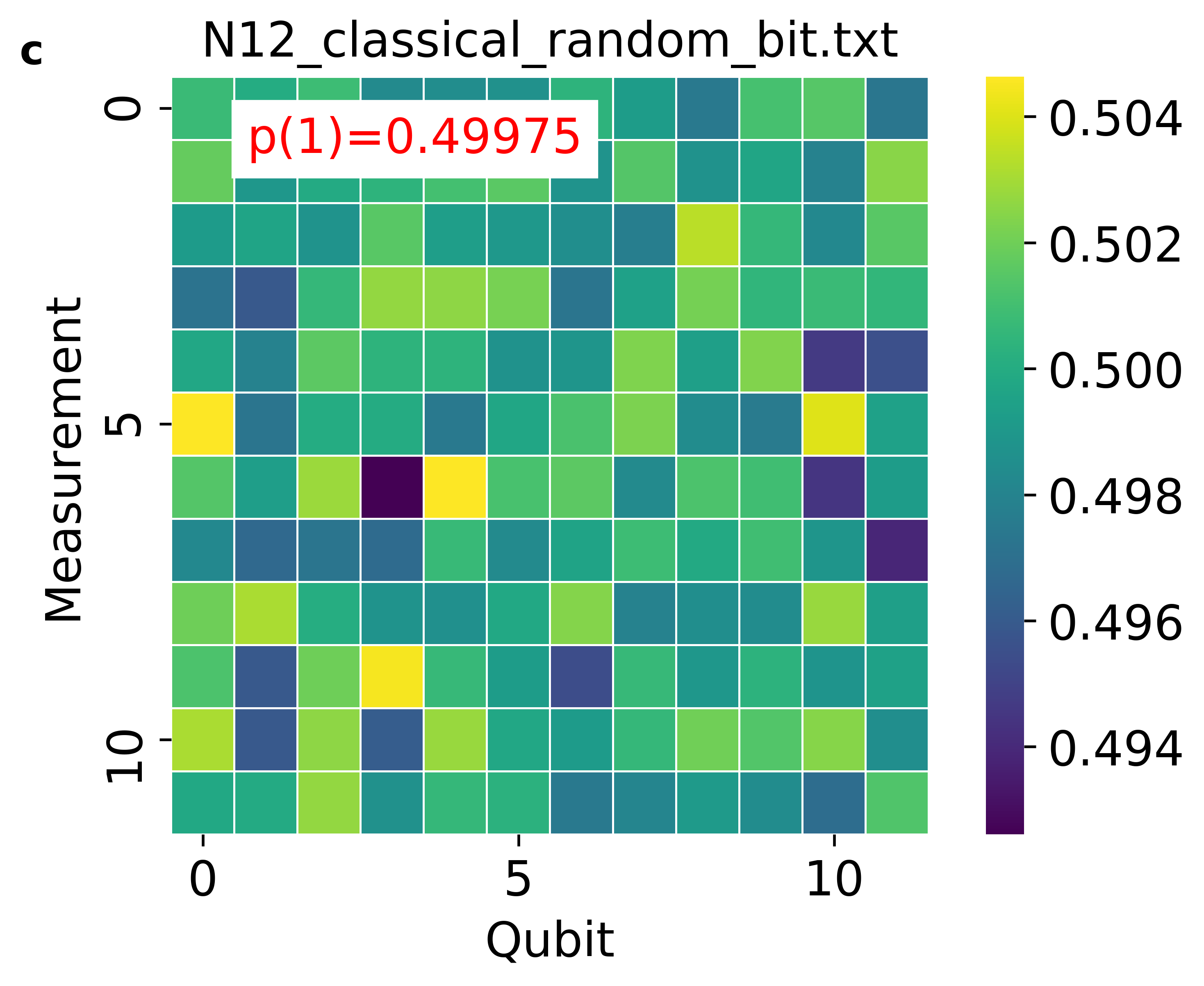}\\[5pt]
\includegraphics[width=0.32\textwidth]{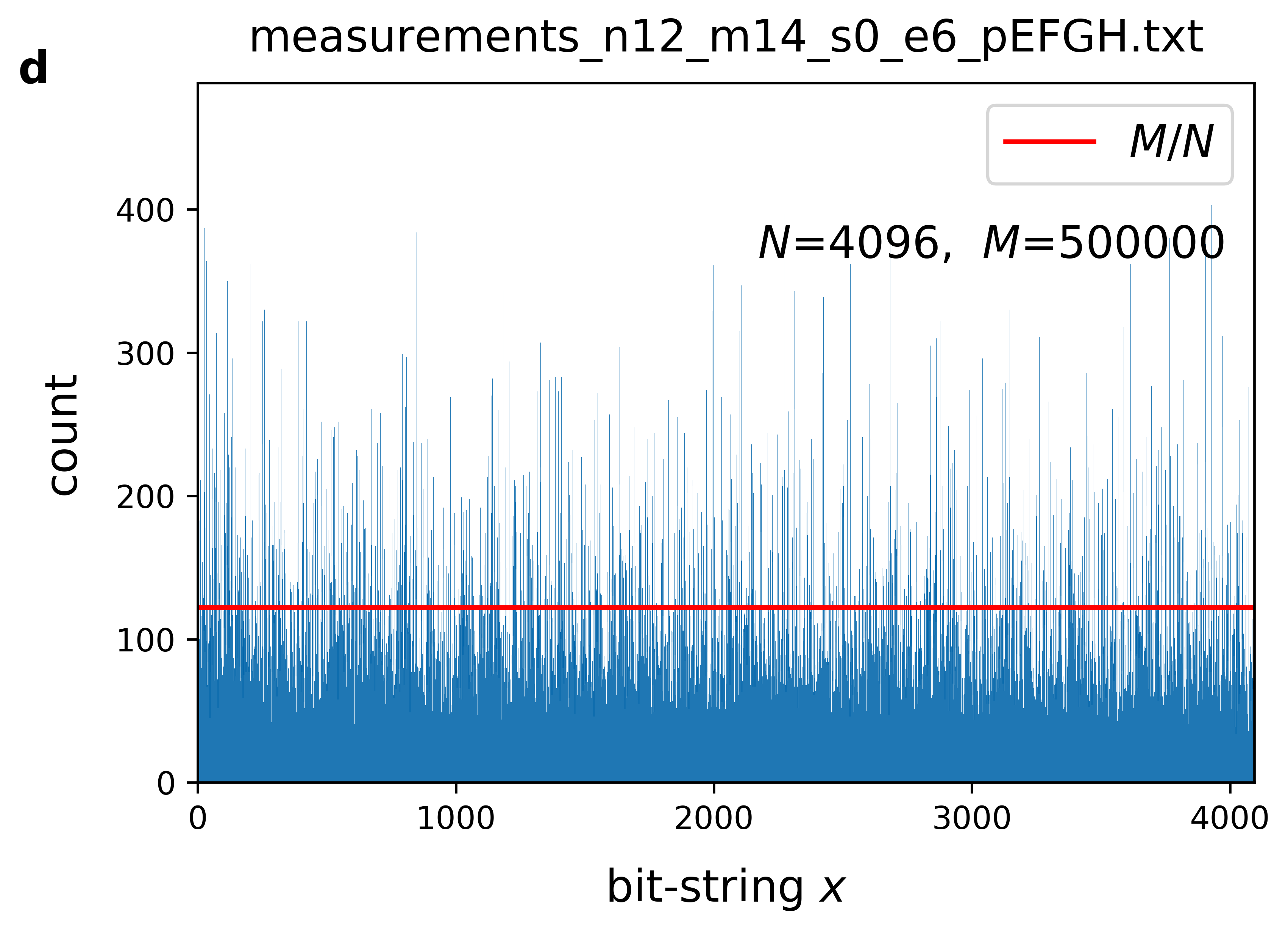}
\includegraphics[width=0.32\textwidth]{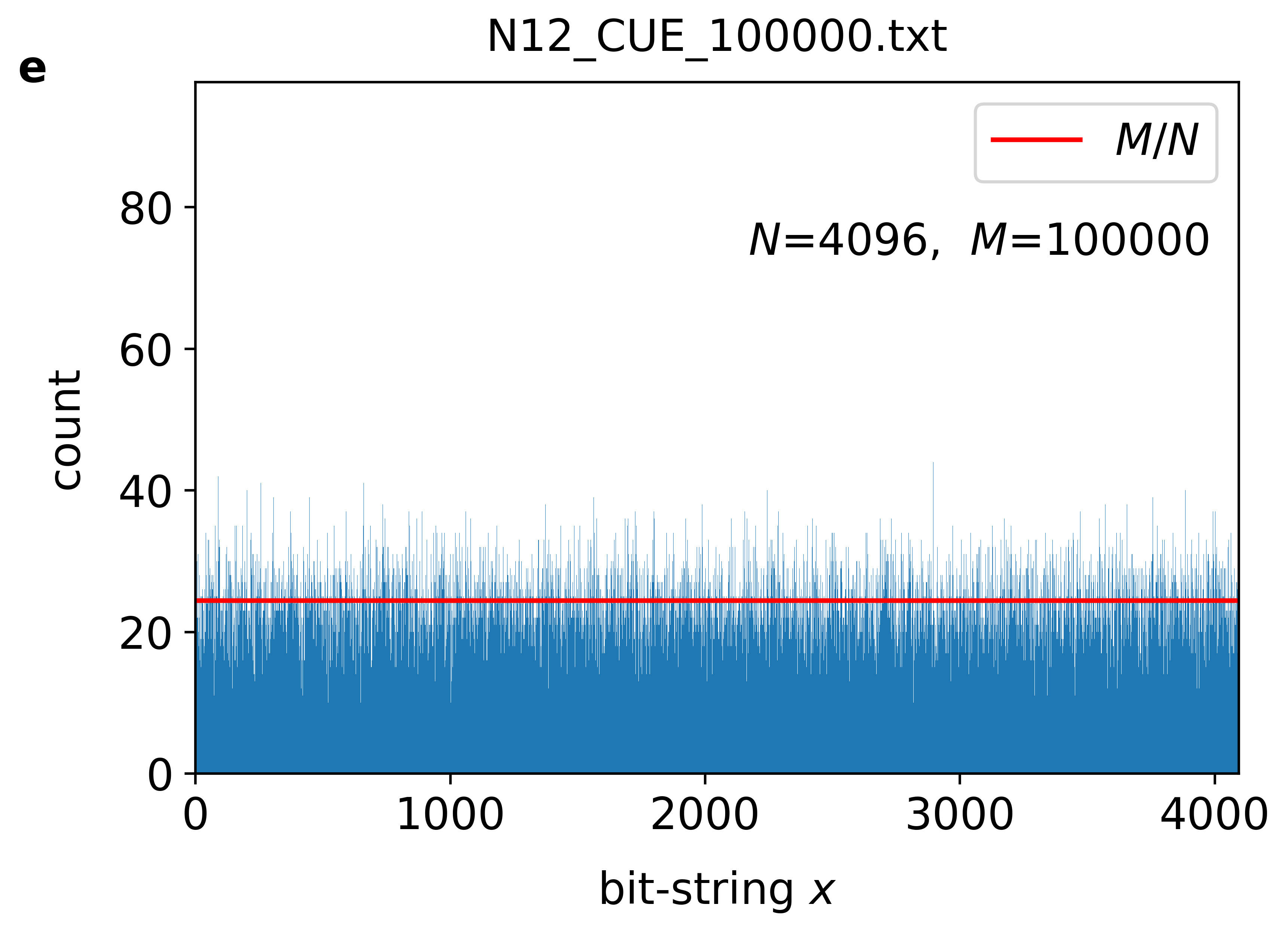}
\includegraphics[width=0.32\textwidth]{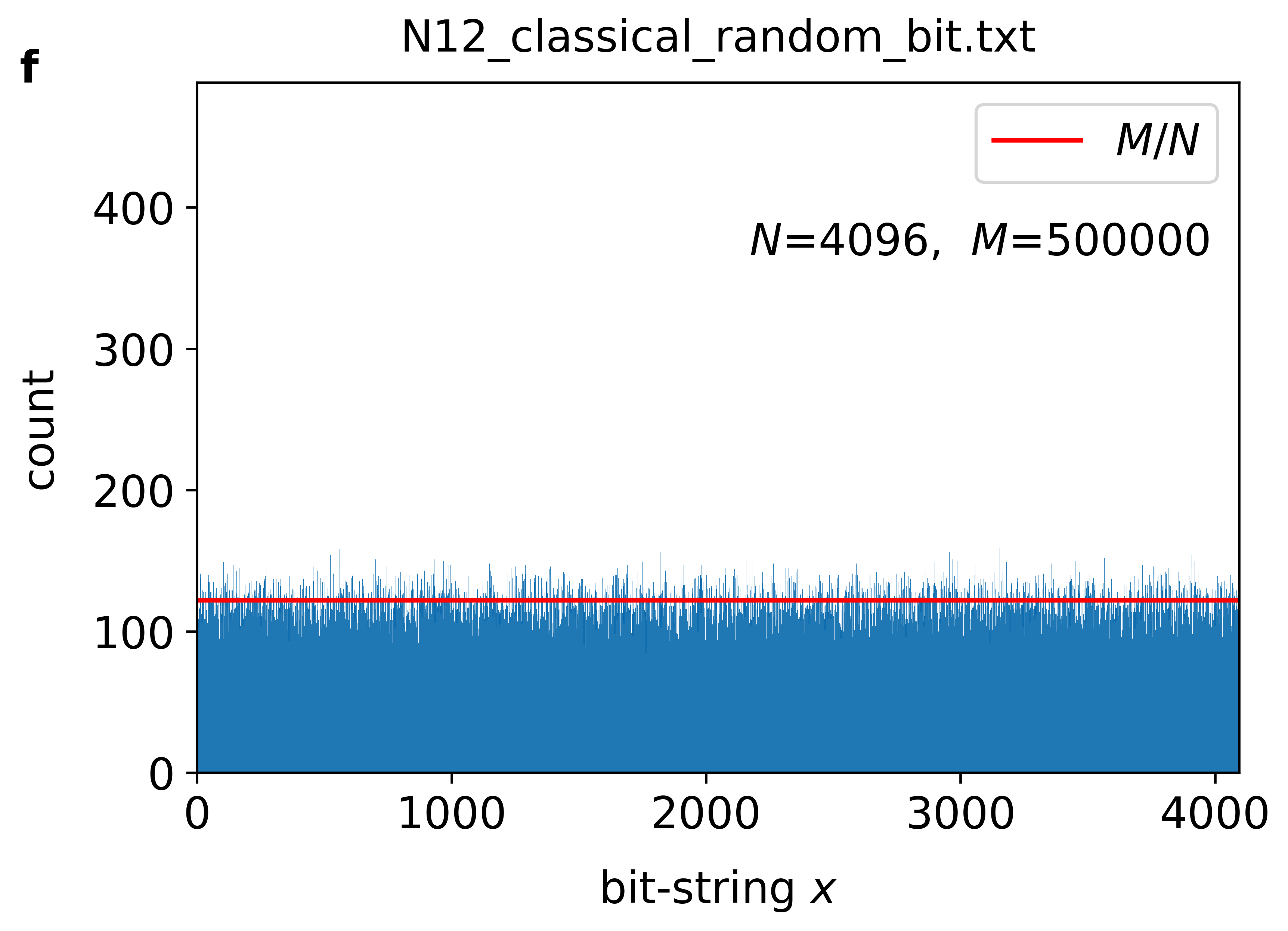}\\[5pt]
\includegraphics[width=0.32\textwidth]{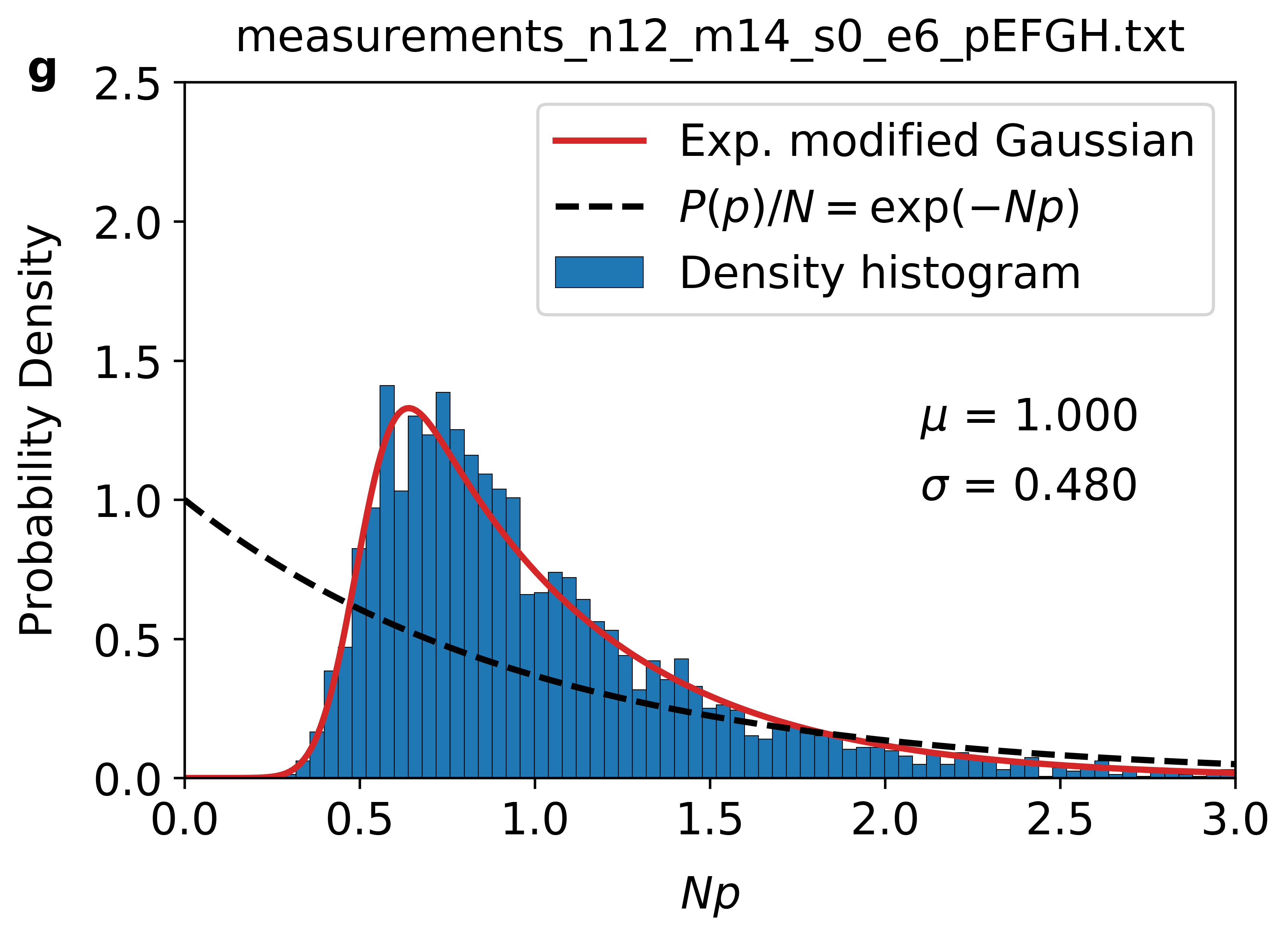}
\includegraphics[width=0.32\textwidth]{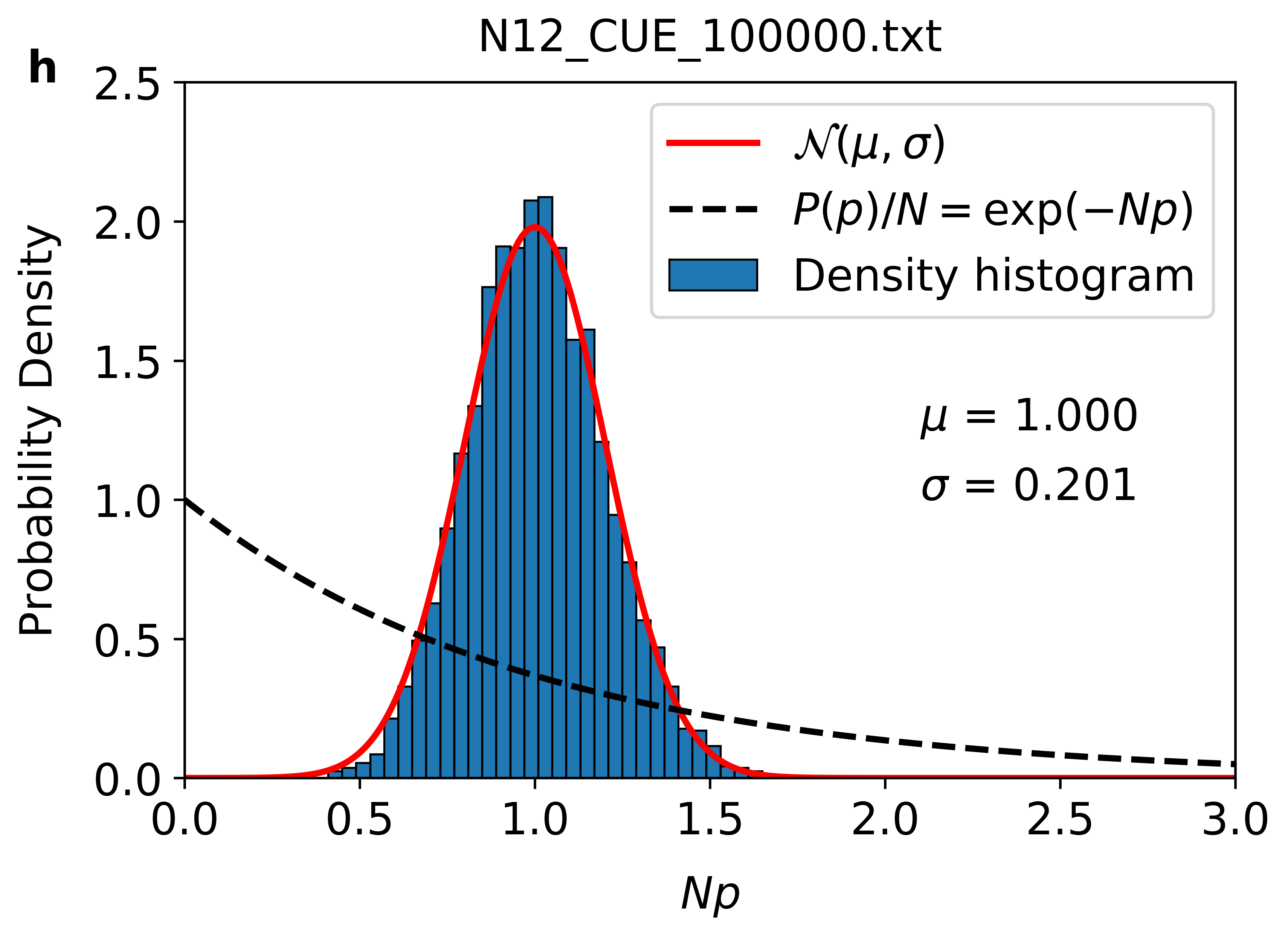}
\includegraphics[width=0.32\textwidth]{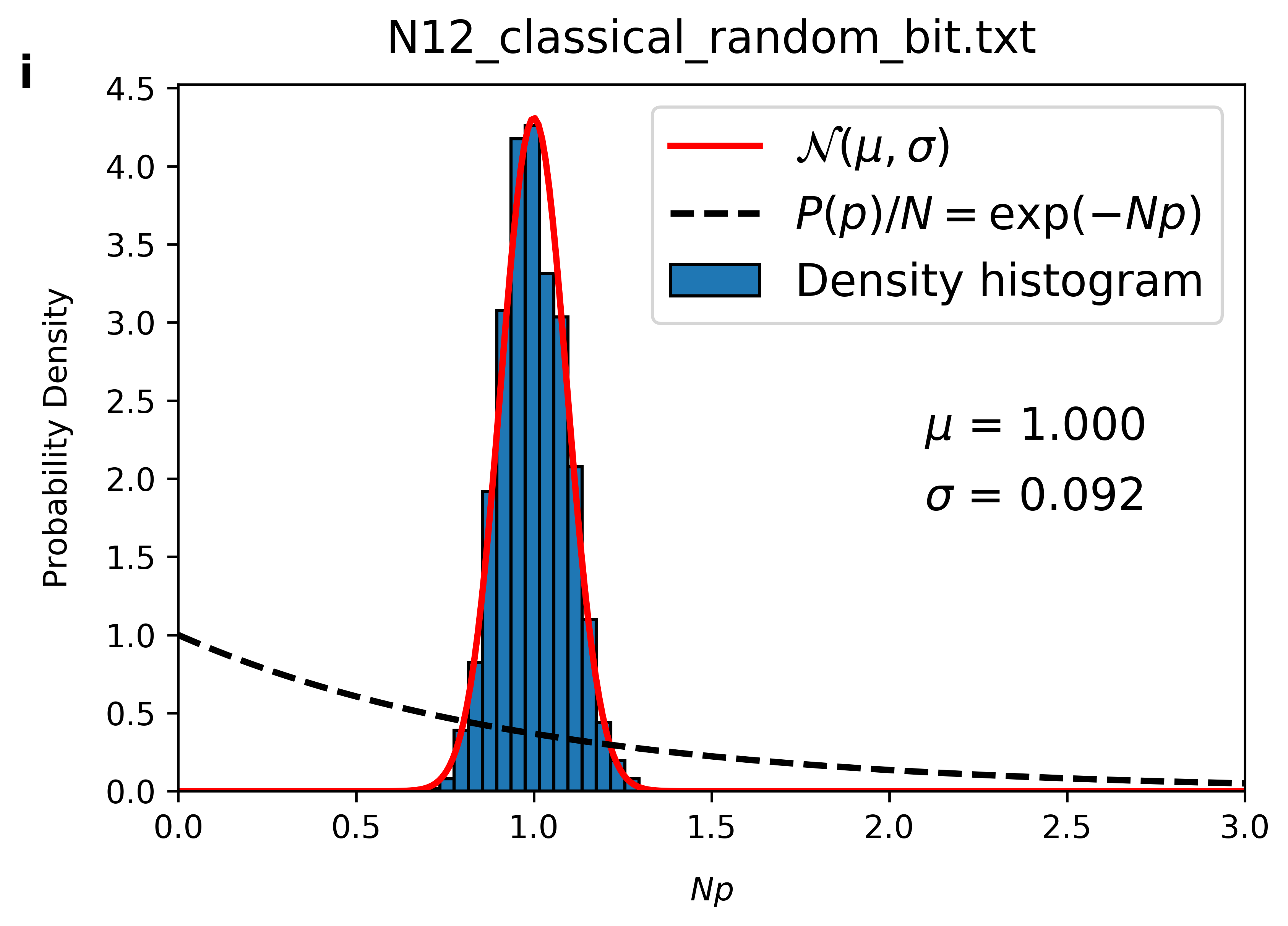}
\caption{\textbf{Three kinds of random-bit strings.} 
The first, the second, and the third columns are figures for Google's random bit-strings circuits 
for $n=12$ qubits and $M=500000$, the CUE random bit-strings for $n=12$ and $M=100000$, classical random 
bit-strings for $n=12$ and $M=500000$, respectively. {\bf a}, {\bf b}, and {\bf c} plot the heat maps 
for random bit strings. $p(1)$ represents the average of getting bit 1. {\bf d}, {\bf e}, and {\bf f} 
are the histograms of bit-strings $x$ as a function of $x$ where $0\le x\le N-1$ 
Here the red lines are the average of $p_x$, $\bar{p} = M/N$.
{\bf g}, {\bf h} and {\bf i} plot the empirical probability density as a function of $Np$ 
constructed from {\bf d}, 
{\bf e}, and {\bf f}. The black lines plot Eq.~(1), $P(p)$. The red lines are fitting curves. 
The exponentially modified Gaussian distribution has three parameters for location, scale, and shape. 
The normal distribution is denoted by ${\cal N}(\mu,\sigma)$ with the mean $\mu$ and the standard deviation $\sigma$.
} \label{Fig1}
\end{figure*}

Google's measurement data sets~\cite{Martinis2021} are the text files of $M\times n$ random-bit arrays labeled by 
five parameters: {\tt n} the number of qubits from 12 to 53, {\tt m} the number of cycles from 12 to 20, 
{\tt s} a seed for the pseudo-random number generator, {\tt e} the elided gate number, and the two types of 
coupler activation patterns (\texttt{EFGH} or \texttt{ABCDCDAB}). The numbers of measurements are $M=500000$, 
$250000$, or $3000000$. One example of Google's data file is \texttt{measurements\_n53\_m20\_s0\_e0\_pABCDCDAB.txt}.

The measurement data, $M\times n$ random bit-strings, are the only information to tell whether the random quantum 
circuit benchmark was implemented properly on Google sycamore qubits. For comparison with Google's random bit-strings, 
we generate (i) random bit-strings sampled from a circular unitary ensemble, called the CUE random bit-strings 
or the Haar measure random bit-strings, and (ii) classical random bit-strings. The CUE random bit-strings are
generated by the QR decomposition algorithm~\cite{Mezzadri2006}. A random matrix $A$ with complex elements sampled
from the normal distribution is factored as $A = QR$, and then $Q$ is a Haar measure unitary matrix. Due to the limited
computing power of a personal computer, the CUE random-bit strings for $n=12$ are generated using Python or Julia 
random matrix library~\cite{Julia_random}. The classical random bit-strings from $n=12$ to $n=53$ are sampled 
using the pseudo-random number generator on a personal computer. The Python scripts generating the CUE and classical 
random bit-strings are attached in Supplementary. 

First, we plot the heat maps of random bit-strings as shown in Fig.~\ref{Fig1}. By slicing a $M\times n$ rectangular 
array of random bit-strings, $D = (x_1,x_2,\dots,x_M)^T$, the ensemble of $n\times n$ square random binary matrices 
$\{X_k = (x_{nk + 1},x_{nk+2},\dots, x_{nk+n})^T\}$ is constructed. Here $x_i = a_1a_2\dots a_n$ is the $i$-th 
row of the random bit array $D$ and $k= 0,1,\dots, M/n -1$. Figs.~\ref{Fig1} (a), (b), and (c) 
show the heat maps of the average density matrix, ${\bf X} = \frac{n}{M}\sum_{k=0}^{M/n} X_k$ with $n=12$ 
for Google, CUE, and classical random bit-strings, respectively. The heat maps for other Google samples 
from $n=12$ to $n=53$ and classical random bit-strings for $n=53$ are shown in Figs.~\ref{Heatmap_Google_EFGH1},
~\ref{Heatmap_Google_EFGH2},~\ref{Heatmap_Google_n53_ABCDCDAB} in Supplementary. 
Surprisingly, all the heat maps of Google's data show the stripe patterns at some qubit indices, while the CUE and 
classical samples do not. As depicted Figs. ~\ref{Heatmap_Google_EFGH1},
~\ref{Heatmap_Google_EFGH2}, and ~\ref{Heatmap_Google_n53_ABCDCDAB} in Supplementary, 
the two coupler activation types, \texttt{EFGH} and \texttt{ABCDCDAB} give rise to the different the bright 
and dark stripe patterns. For $n=53$ and $\texttt{ABCDCDAB}$ activation, the stripe patterns become more clear 
as the cycle number $m$ increases. The total number of bit 1 of $D$ is counted to calculate the average of 
finding bit 1, $p(1)$. As shown in Figs.~\ref{Fig1} (a), (b), and (c), Google's random bit-strings 
show the vale $p(1) \approx 0.486$, that is less than the expected value $0.5$, while the CUE and classical 
random bit-strings have $p(1)\approx 0.499$ very close to 0.5. As shown in Figs.~\ref{Heatmap_Google_EFGH1},
~\ref{Heatmap_Google_EFGH2},~\ref{Heatmap_Google_n53_ABCDCDAB}, $p(1)$ for Google's random bit strings 
ranges from 0.483 to 0.489. The non-randomness of Google's random bit-strings can also be checked with 
a random number test, as well as the stripe patterns. We perform the NIST statistical random number 
tests~\cite{NIST2010,Ang2019}. As shown in Table~\ref{Random_number_test} in Supplementary, Google's quantum random 
bit-strings fail to pass some NIST random number tests, while the CUE and classical random bit strings pass.
The failure of the NIST frequency test means Google's data has too many 0's. Also Google's data fail
to pass some NIST tests. Thus, we demonstrate that Google's random bit-strings are not truly random 
in contrast to the CUE or classical random bit-strings.

Next, we calculate the empirical distributions $p_x = b_x/M$ for $n=12$, where $b_x$ counts the number of 
bit-strings with the value $x$ and $0\le x \le N-1$ in decimal notation. As illustrated in Figs.~\ref{Fig1} 
(d), (e), and (f), Google's random bit-strings fluctuate widely and the classical random bit-strings
fluctuate a little. Using the empirical distribution $p(x)$, one can construct the three empirical probability 
distributions $P_{\rm Google}(p)$, $P_{\rm CUE}(p)$, and $Q_{\rm cl}(p)$ to see they follow 
Eq.~(\ref{dist_eigenvector_CUE}), the eigenvector distribution of the circular unitary ensemble, as shown
in Fig.~\ref{Fig1} (g), (h), and (i). All three distributions are different from the eigenvector 
distribution $P(p)$ of the CUE, Eq.~(\ref{dist_eigenvector_CUE}). We leave the study of these discrepancies
as an open question. 


\begin{figure}[t]
\includegraphics[width=0.48\textwidth]{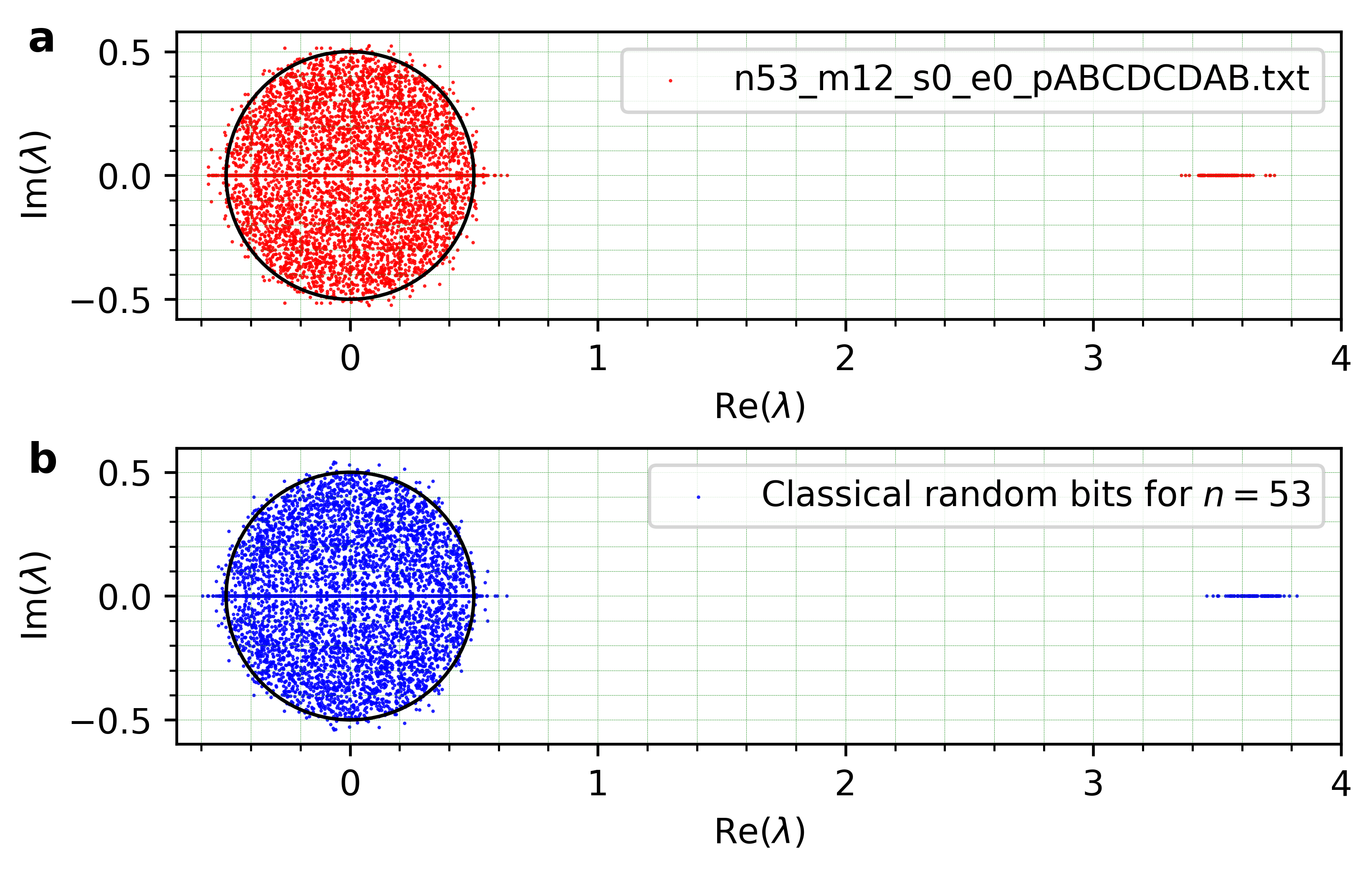}
\caption{\textbf{Circle law of random matrices of random bit-strings.} 
The eigenvalue distributions of the $n\times n$ random bit matrices $\{X_k\}$ 
are shown for Google's random bit-strings {\bf a} and for the classical random bit-strings {\bf b} 
for $n=53$. Only eigenvalues of 100 $X_k$ samples are shown. 
The black circle is known as the Girko circle of the non-Hermitian Ginibre ensemble. 
The outliers far away from the circle are the real eigenvalues located between 3 and 4.}
\label{circle_law}
\end{figure}

The difference between Google random bit-strings and the CUE or classical random bit-strings is further illustrated 
using the random matrix theory of the random binary matrices. The collection $\{X_k\}$ of $n\times n$ matrices 
$X_k$ of random bit-strings can be regarded as a real Ginibre ensemble. It is well known that for random matrices with 
identically-and-independently matrix elements and zero mean, the distribution of complex eigenvalues of random matrices 
follows the Girko circular law~\cite{Girko1984,Tao2015}. However, random matrices formed by random bit strings here
have the matrix element $x_{ij} \in  \{0, 1\}$, so the mean of a matrix $X$ is not zero.
If $x_{ij}$ are sampled identically and independently from the Bernoulli distribution, the mean of $X$ could be $1/2$.
We are interested in whether the mean of $X$ generated by quantum random circuits is identical to 1/2 or not.

The distribution of the complex eigenvalues of random matrices $\{\frac{1}{\sqrt{n}}X^{(k)}\}$ with $k=1,...,100$ are 
plotted in Fig.~\ref{circle_law}. Most eigenvalues of both Google and classical random matrices are distributed 
inside of the circle with radius 1/2 and some outliers with large real eigenvalues are located outside the circle. 
As shown in Fig.~\ref{circle_law} closely, the positions of outliers with large real eigenvalues of  
Google's random bit-strings are different from that of the classical random bit-strings.
The radius 1/2 and the outliers can be explained as follows.
The matrix $X$ with nonzero mean can be transformed to $Z$ with zero mean by 
\begin{align}
Z = 2X - J \,,
\label{Eq_trans}
\end{align}
where $J$ is an all-one matrix. 
The ensemble of $n\times n$ real random square matrices $(Z_k)$ with the matrix elements $z_{ij}$ 
sampled identically and independently from the Bernoulli distribution with zero mean and unit variance. 
The complex eigenvalues of $\frac{1}{\sqrt{n}}Z$ are distributed uniformly in the unit circle.  
So the radius of the circle of the eigenvalue distribution of $X$ is 1/2.
The Saturn-ring effect along the real line and the outliers shown in Fig.~\ref{circle_law} are due to the fact 
that $X$ and $J$ are non-commutative. 


\begin{figure}[t]
\includegraphics[width=0.45\textwidth]{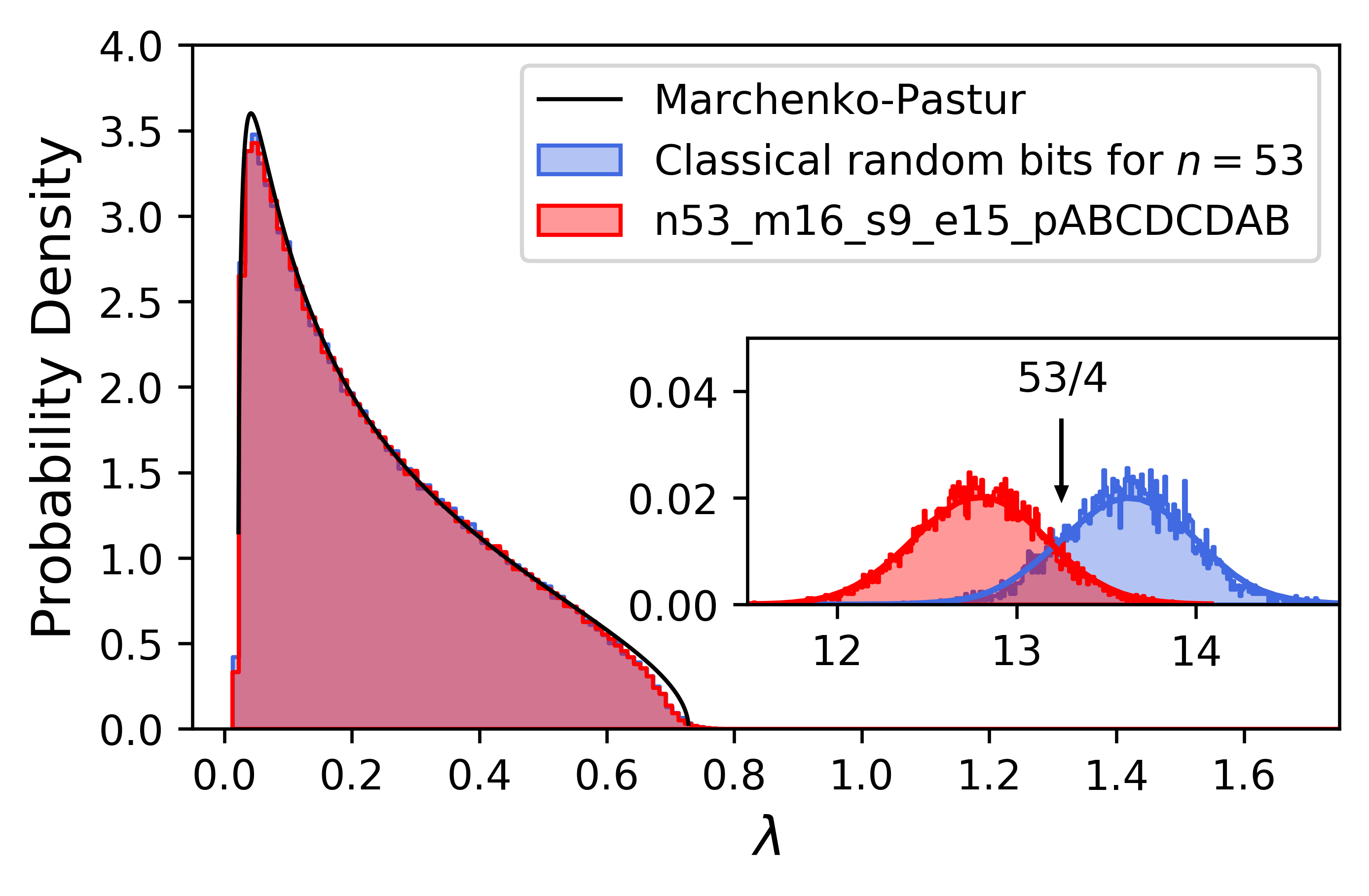}
\caption{\textbf{Marchenko-Pastur distribution of the Wishart ensemble of random bit-strings.} The Marchenko-Pastur 
distribution of eigenvalues of Wishart ensembles 
for Google's random bit strings and classical random bit strings.}
\label{MP_dist}
\end{figure}

Let us slice the $M\times n$ random bit-string array $D$ into $p\times n$ rectangular binary matrices $X$ 
where $p>n$. Then, the collection of $n\times n$ symmetric matrices $W = \frac{1}{p} X^t\cdot X$ is called
the Wishart ensemble. It is known that if the elements of $X$ are sampled identically and independently 
from the normal distribution ${\cal N}(\mu,\sigma)$ with zero mean $\mu =0$ and the variance $\sigma^2$, 
the distribution of real eigenvalues of $W$ is given by the Marchenko-Pastur distribution~\cite{MarPas67}
\begin{align}
\rho(\lambda) = \frac{1}{2\pi\gamma\sigma^2} \frac{\sqrt{(\lambda_+ -\lambda)(\lambda -\lambda_{-})}}{\lambda}\,,
\end{align}
where $\lambda_{\pm} = \sigma^2(1 \pm\sqrt{\gamma})^2$ are the upper and lower bounds and 
$\gamma = n/p$ is the rectangular ratio.
Here we take $p=1/2$. Fig.~\ref{MP_dist} plots the Marchenko-Pastur distributions of 
Google's and classical random-bit strings for $n=53$. As shown in Fig.~\ref{MP_dist},  
the outliers outside the Marchenko-Pastur distribution distinguish Google's random bit-strings from
the classical random bit-strings. With Eq.~(\ref{Eq_trans}), $W$ can be expressed as 
\begin{align}
W = \frac{1}{4p}\left(Z^t\cdot Z+ Z^t\cdot J + J^t\cdot Z + J^tJ\right) \,. 
\label{Wishart}
\end{align}
Here the first term of Eq.~(\ref{Wishart}) is be written as $\frac{1}{p} \frac{Z^t}{2}\cdot\frac{Z}{2}$, so 
$\frac{Z}{2}$ has zero mean and variance $\sigma^2 = 1/4$ while the variance of $X$ is $\sigma^2 = 1/2$.
This gives the upper and lower bounds, $\lambda_{+} = 0.728$ and $\lambda_{-}= 0.021$, respectively.
The last term of Eq.~(\ref{Wishart}) becomes $\frac{1}{4p}(J^t)_{n\times p}\cdot (J)_{p\times n} 
= 1/4 (J)_{n\times n}$ where an all-one matrix $J$ has the eigenvalue 0 and $n$. So
the outliers are located around $53/4$.

\smallskip
\noindent
\paragraph*{\bf How far away are they? -- Wasserstein Distances\\} 

\begin{figure}[t!]
\includegraphics[width=0.48\textwidth]{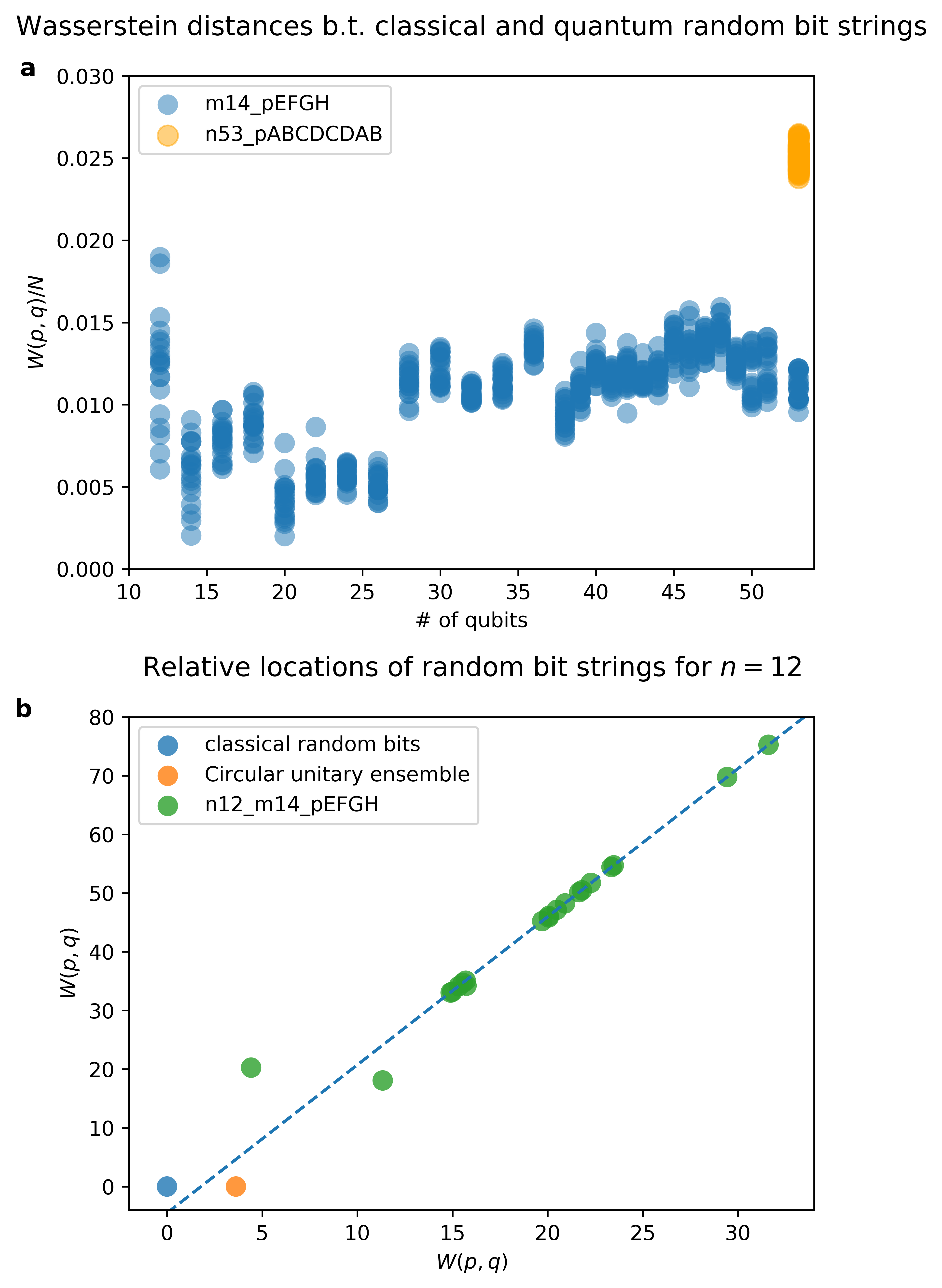}
\caption{\textbf{Wasserstein Distances.} {\bf a} The Wasserstein distances, divided by $N$, between Google's 
quantum random bit-strings and the classical random bit-strings are plotted as a function of the number of qubits, 
$n$. The blue and orange dots represent the activation patterns \texttt{EFGH} and \texttt{ABCDCDAB}, respectively. 
{\bf b} For $n=12$, All pairs of Wasserstein distances among Google, CUE, and classical random bit samples are 
calculated and their relative locations are plotted.
For each sample, $M=500,000$ is taken.}
\label{Fig4}
\end{figure}

Up to now, we have shown (i) the non-randomness of Google's random bit-strings using the heat maps and (ii) 
the random matrix theory of random bit-strings can distinguish Google's data from the classical random samples.
The final question we would like to address is how much Google random bit-strings are different 
from the CUE or classical random bit-strings. In Google's experiment, the cross entropy fidelity was used to 
measure how the real distribution is close to the ideal distribution. The disadvantage of the cross entropy 
is that it is not symmetric and gives rise to zero or diverge if there is no overlap between two distributions. 
To overcome these, we employ the Wasserstein distance of order 1, $W(p,q)$ between two discrete probability 
distributions, $p_i$ and $q_i$,
\begin{align}
W(p,q) =  \inf_{\lambda_{ij},i,j} \sum_{i,j} \lambda_{i,j} |x_i - y_j|\,,
\end{align}
where $\lambda_{ij}$ is the joint probability of $x_i$ and $y_j$ such that $\sum_{i} \lambda_{i,j} = q_i$, 
$\sum_{j} \lambda_{ij} = p_i$, and $\lambda_{ij}\ge 0$. Given two samples, $\{x_1,x_2,\dots,x_M\}$ and 
$\{y_1,y_2,\dots,y_M\}$, $W(p,q)$ can be obtained directly without calculating the empirical distributions 
$p$ and $q$. We use the Python optimal transport library~\cite{flamary2021pot} for calculating 
the Wasserstein distance between two samples. Fig.~\ref{Fig4} (a) plots the Wasserstein distances, normalized 
by $N$, between Google and classical random bit-strings as a function of $n$. For $n=53$, Google samples with 
the activation pattern with {\tt EFGH} are closer to the classical random bits than that with {\tt ABCDCDAB}. 
For $n=12$, we calculate the Wasserstein distance among all pairs of Google, CUE, and classical random 
bit-strings so their relative locations in the 2-dimension are displayed in Fig.~\ref{Fig4} (b). 
It shows that Google random samples are farther away from the CUE random bit sample than the classical random 
bit sample. Also, all Google samples except 2 samples are fit to a straight line passing between the CUE
and classical random bit samples.

\smallskip
\paragraph*{\bf Summary\\}
In conclusion, we analyzed the randomness of Google's quantum random bit-strings generated by random quantum 
circuits. It is found that the heat maps of Google's data have the stripe patterns at specific qubit sites and 
contain more bit 0 than bit 1. This led to Google's data failing to pass the NIST random number tests.
These non-randomness signatures of Google bit strings do not occur in the Haar-measure random sample or in 
the classical random bit-strings. The non-randomness of Google's data is caused by the errors of 
quantum gates or by the readout errors. The random matrices of random bit-strings distinguish Google data
from the classical random bit-strings. The calculation of the Wasserstein distance shows that Google random bits 
are farther away from the Haar measure sample than the classical random bit sample. The Wasserstein distance between 
two samples would be a simple and powerful tool of measuring the fidelity of quantum gates.  
The two activation patterns of Google's Sycamore qubits give rise to quite different results in the heat map of 
the bit strings and in the Wasserstein distances. 
Our findngs imply that the random matrix analysis and the Wasserstein distance may be used as benchmark tools
to measure the performance of intermediate scale quantum computers. The linear cross-entropy used in Google's supremacy 
experiment needs the classical simulation to calculate the probability $p(x)$ of finding a bit-string $x$.
As the number of qubits increases, thise classical simulation is very difficult. So one faces the dilemma of
how to veryfy the performance of quantum computers using the linear cross-entropy~\cite{Eisert2020}.
The calculation of the Wasserstein distance requires only two data sets of bit-strings and do not need
to estimate probability distributions from the two data sets. Classical random bit-strings for one hundred qubits 
can be easily generated and the Wasserstein distance between the classical random bit-strings and
the random bit-strings of random quantum circuits could be calculated.


\paragraph*{\bf Acknowledgements}
We would like to thank Seth Lloyd, Scott Aaronson, and Sergio Boixo for reading and commenting on our draft.
{\bf Funding:}
This material is based upon work supported by the U.S. Department of Energy, Office of Science, 
National Quantum Information Science Research Centers.
We also acknowledge the National Science Foundation under award number 1955907. 
{\bf Author contributions:} 
S.K. led and coordinated the project. S.O. conceived the research idea and performed numerical calculations, 
All authors discussed the results and wrote the manuscript.
{\bf Competing interests:} The authors declare no competing interests.
{\bf Data and material availability:} All data and codes are provided in the main text or the supplementary 
materials.
\bigskip

\bibliography{Main_Q_supremacy.bib}

\pagebreak
\clearpage

\begin{widetext}
\section*{Supplementary Materials for ``Non-Randomness of Google's Quantum Supremacy Benchmark''}

\renewcommand{\thefigure}{S\arabic{figure}}
\renewcommand{\thetable}{S\arabic{table}}
\setcounter{figure}{0}
\setcounter{table}{0}
\NewDocumentCommand{\rot}{O{45} O{1em} m}{\makebox[#2][l]{\rotatebox{#1}{#3}}}%

\subsection{\bf Preparation of the data and the script files}
The three kinds of random bit strings used in this study are prepared as follows.
\begin{itemize}
\item Google's data for quantum supremacy benchmark test~\cite{Martinis2021} is downloaded from the Dryad Digital Repository,
\verb|https://datadryad.org/stash/dataset/doi:10.5061/dryad.k6t1rj8|.
\item The QR algorithm is used to perform the Haar measure sampling of unitary operators from $U(N)$. The Python script and the 
Julia script, {\tt Haar.py} and {\tt Haar.jl} were written. Due to the limit of the computational power of the PC 
(Intel Core i7-4790 CPU @ 3.60 GHz and 24 GB memory), it took about 7 days to sample 60000 random unitary operators using 
the QR algorithm for $n=12$.
\item The classical random bit strings are generated using a simple Python script, {\tt Randbit.py}. The classical random bit
array of the size $n=53$ by $M=500000$ is produced quickly within a few minutes on the PC. 
\end{itemize}

\subsection{Eigenvalue distributions of the Haar measure samples}
Fig.~\ref{Eigenvalue_CUE} shows the eigenvalue distributions of $M=1000$ Haar measure samples of $U(2^{12})$ generated by the QR
algorithm.
\begin{figure}[h]
\includegraphics[width=0.25\textwidth]{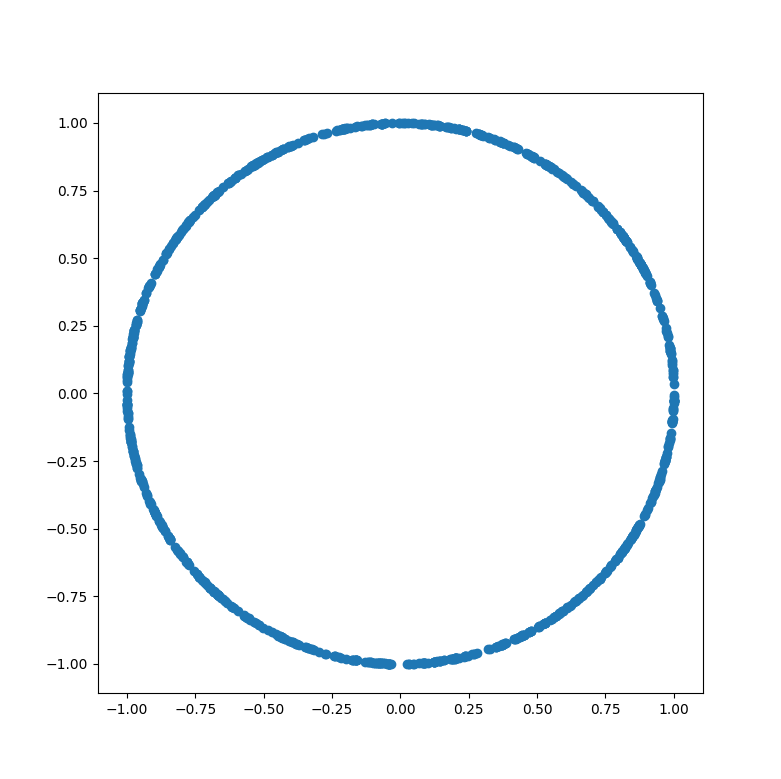}
\caption{Eigenvalue distribution of circular unitary ensembles}
\label{Eigenvalue_CUE}
\end{figure}

\subsection{Heat maps of Google random bit-strings}

Figs.~\ref{Heatmap_Google_EFGH1} and ~\ref{Heatmap_Google_EFGH2} 
show the heat maps of Google's data from $n=12$ to $n=53$ with the activation pattern {\tt EGGH}. 
Figs.~\ref{Heatmap_Google_n53_ABCDCDAB} depicts the heat maps of Google's data for $n=53$, the cycles $m=12,14,16,20$, and
the activation pattern {\tt ABCDCDAB}. The pictures are plotted by running the Python script, {\tt Heatmap.py}, as follows 

\pagebreak

\begin{figure*}[h]
\includegraphics[width=0.3\textwidth]{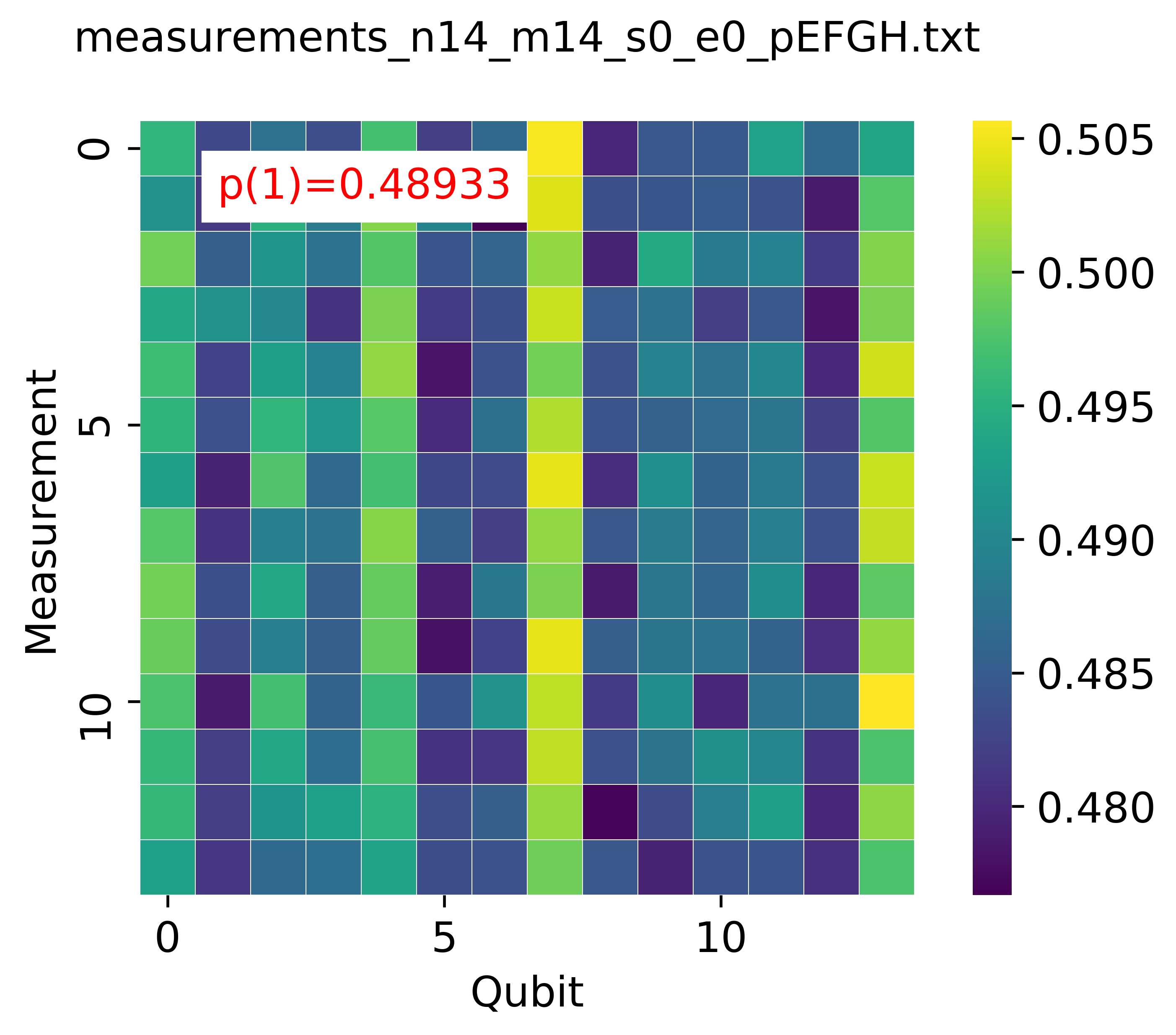}
\includegraphics[width=0.3\textwidth]{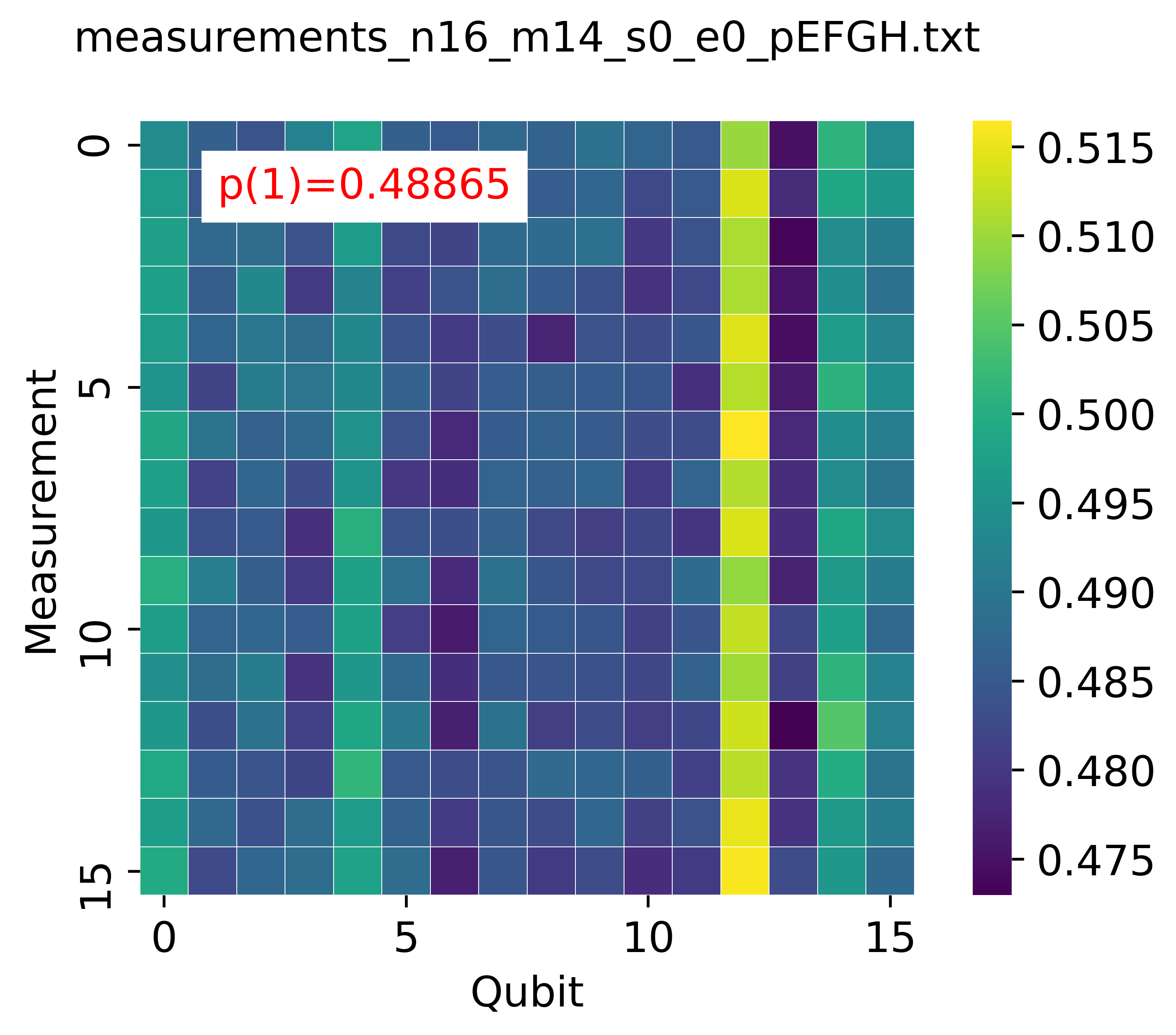}
\includegraphics[width=0.3\textwidth]{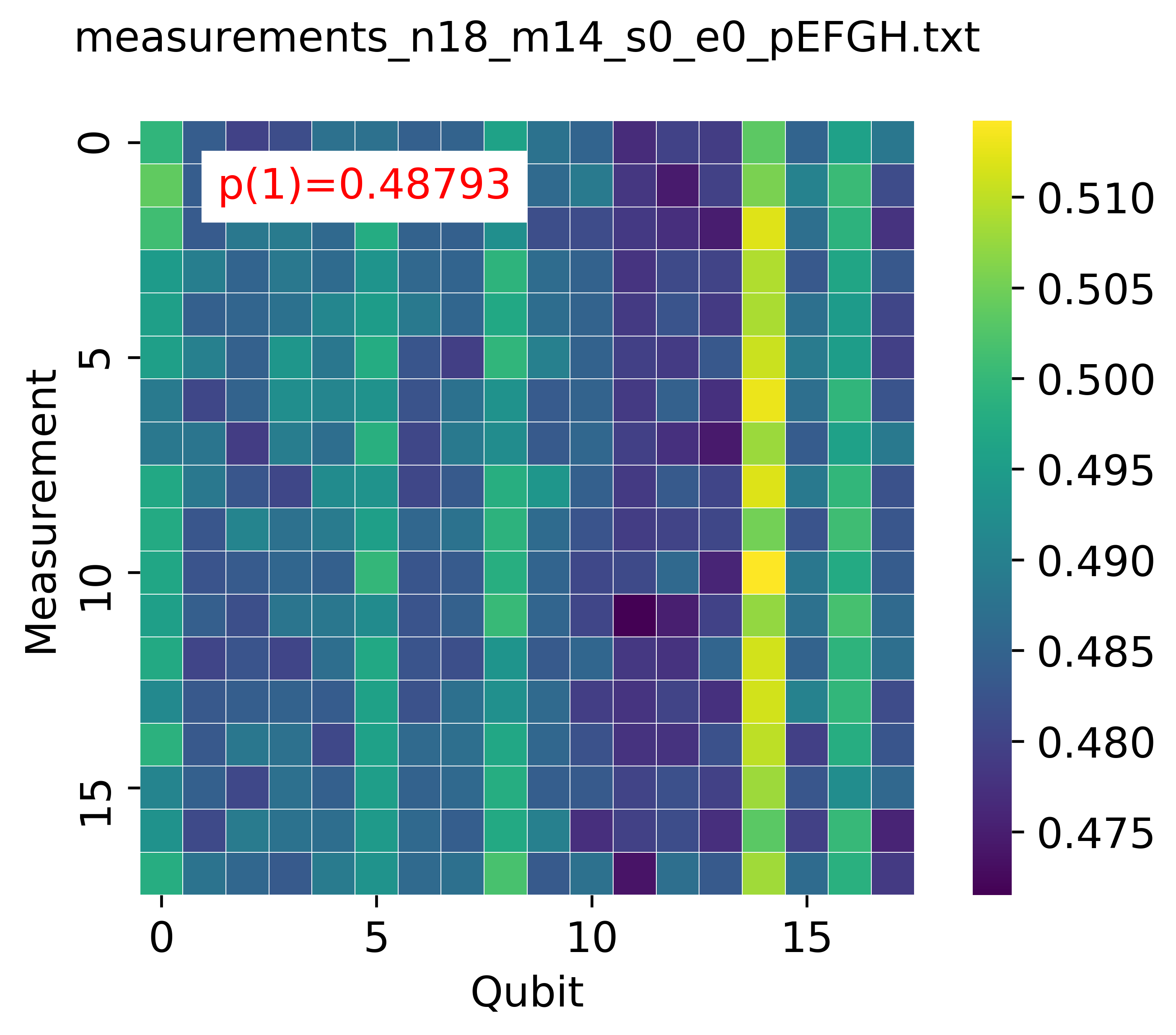}
\includegraphics[width=0.3\textwidth]{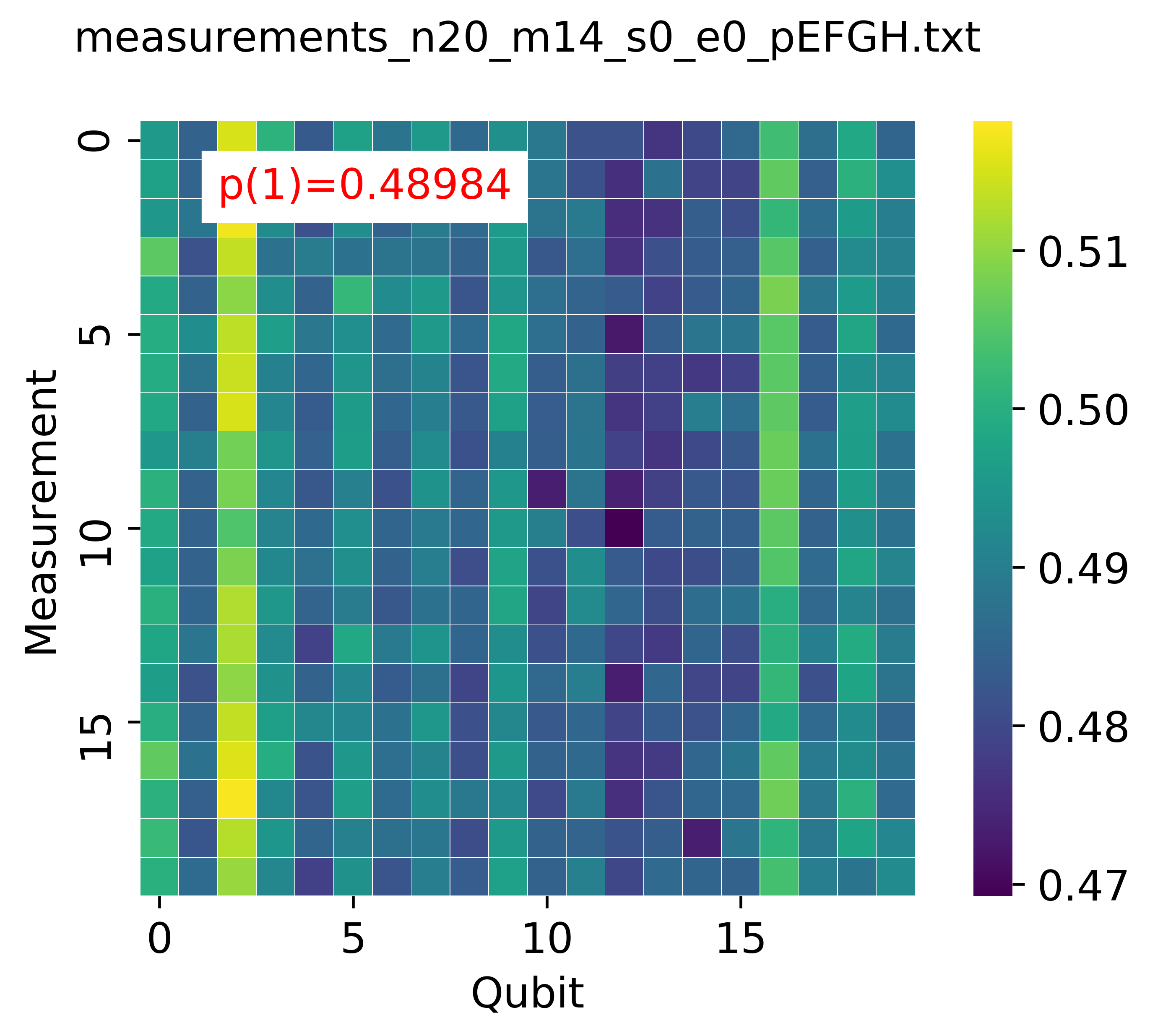}
\includegraphics[width=0.3\textwidth]{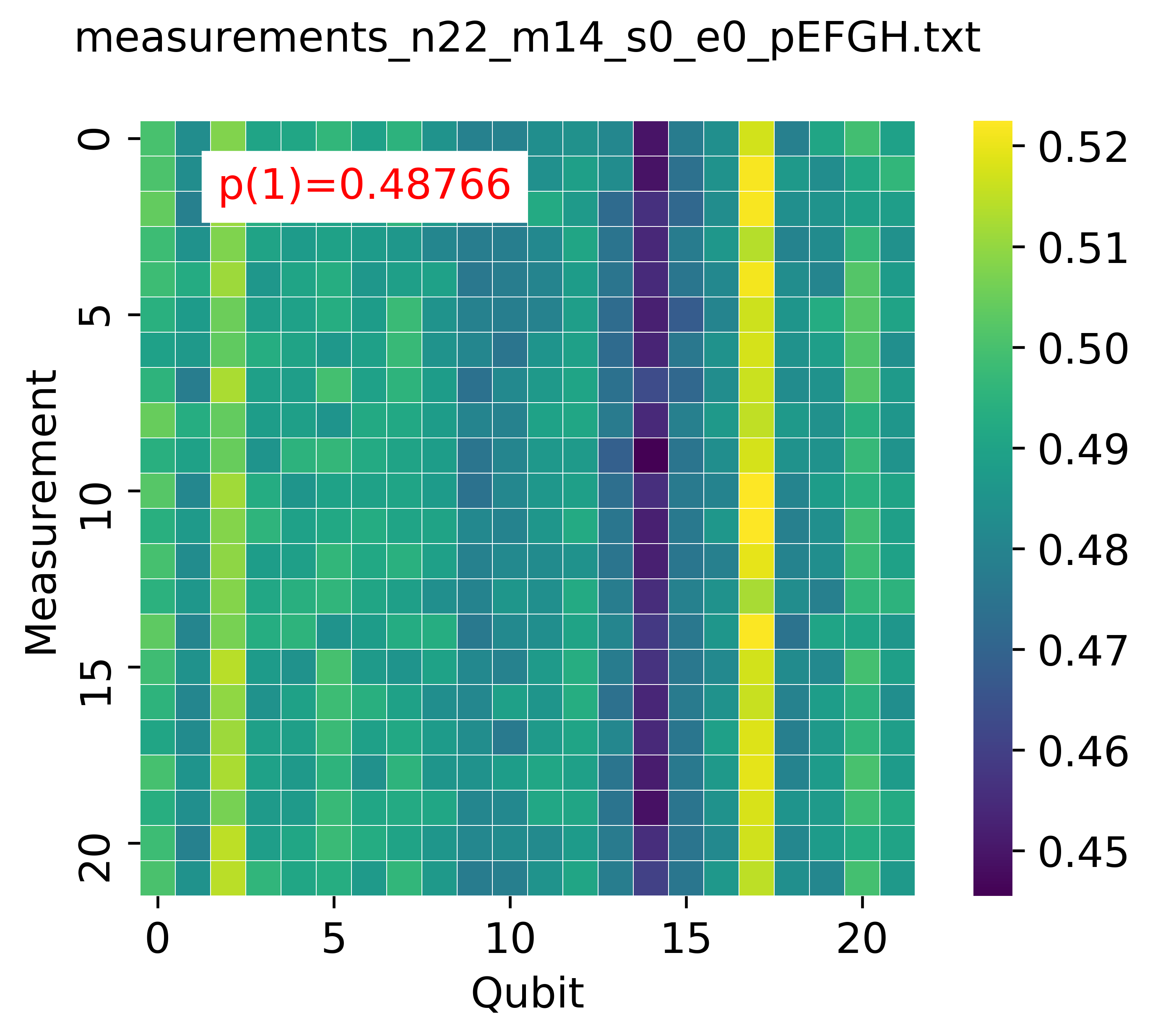}
\includegraphics[width=0.3\textwidth]{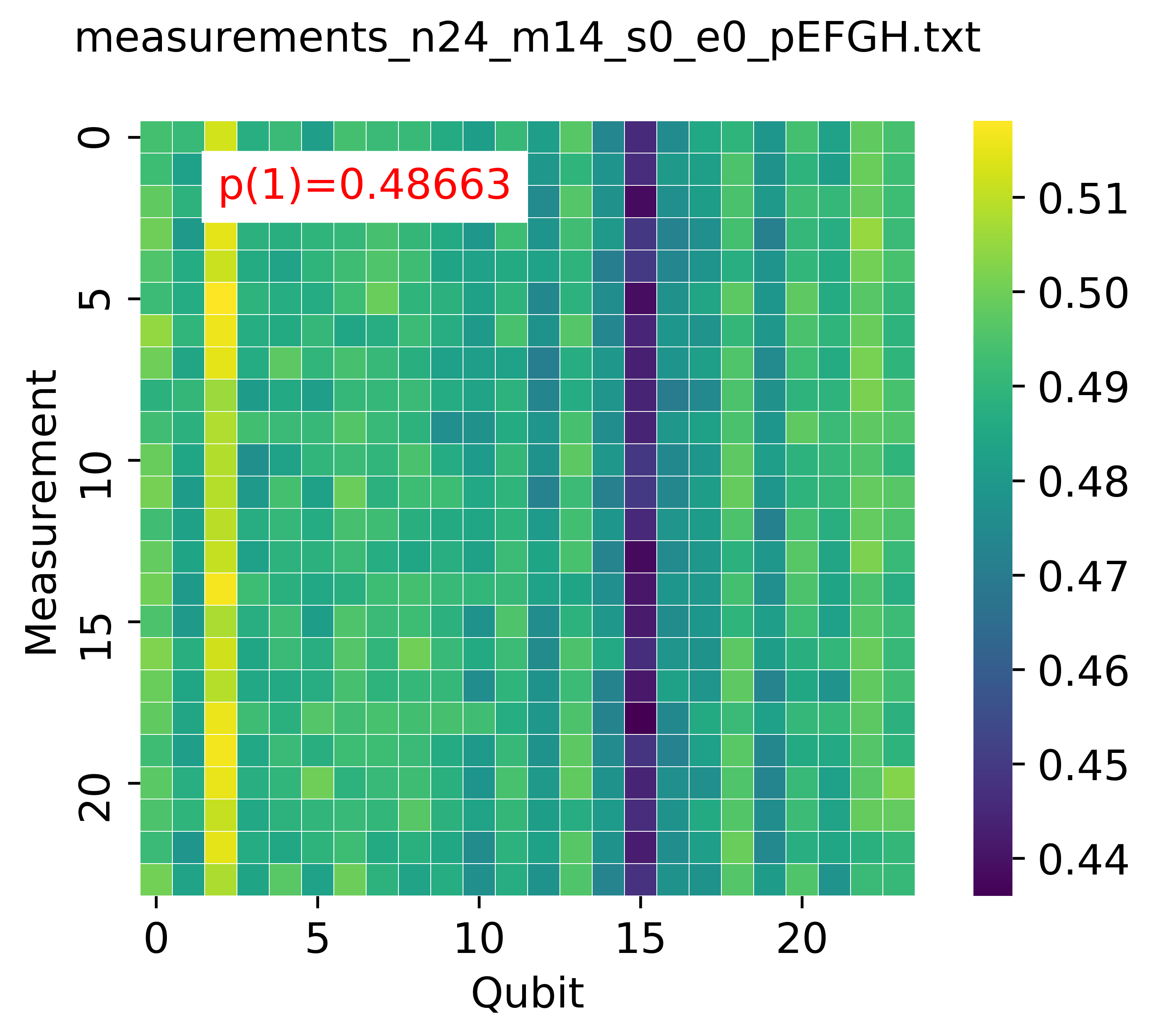}
\includegraphics[width=0.3\textwidth]{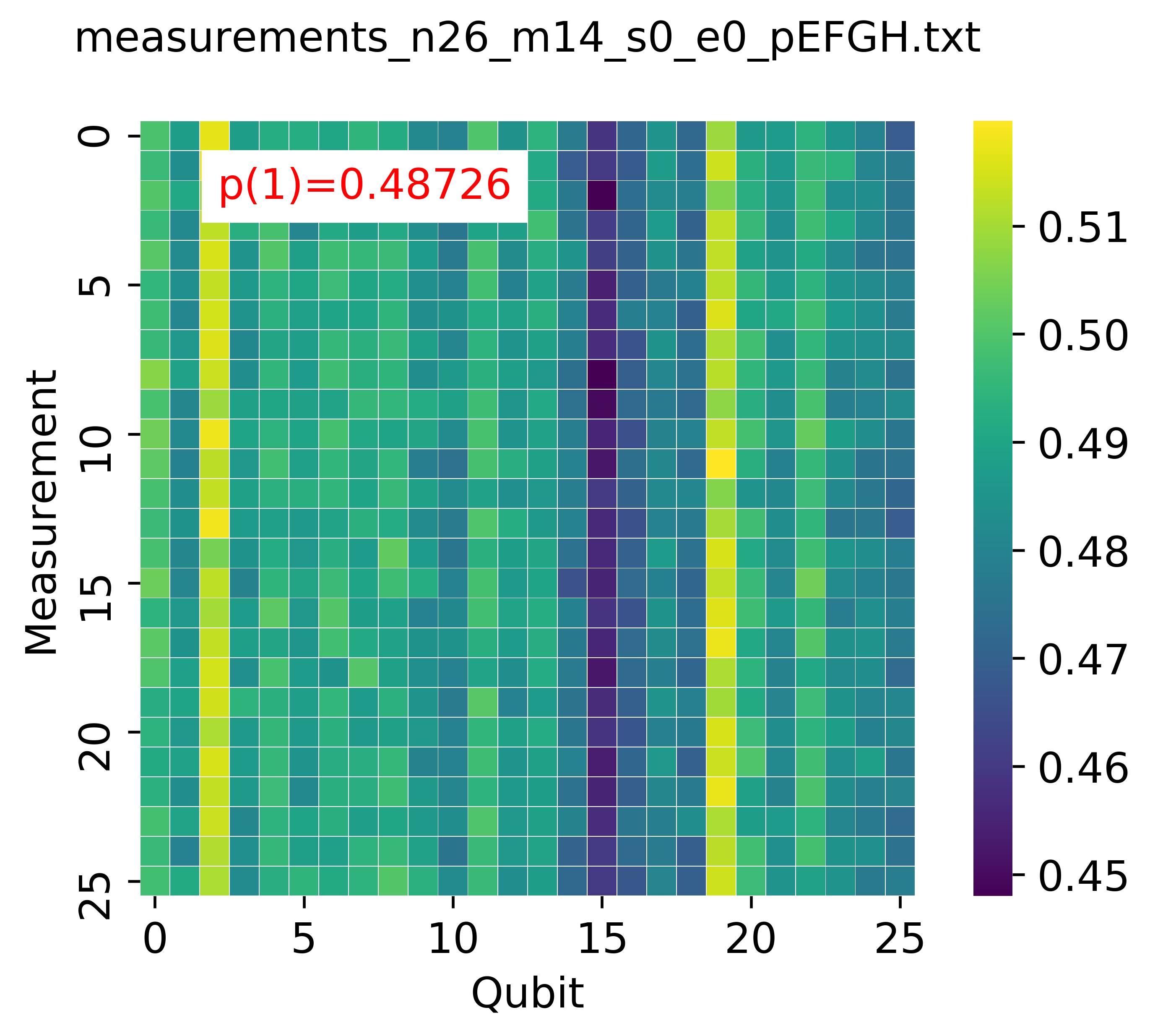}
\includegraphics[width=0.3\textwidth]{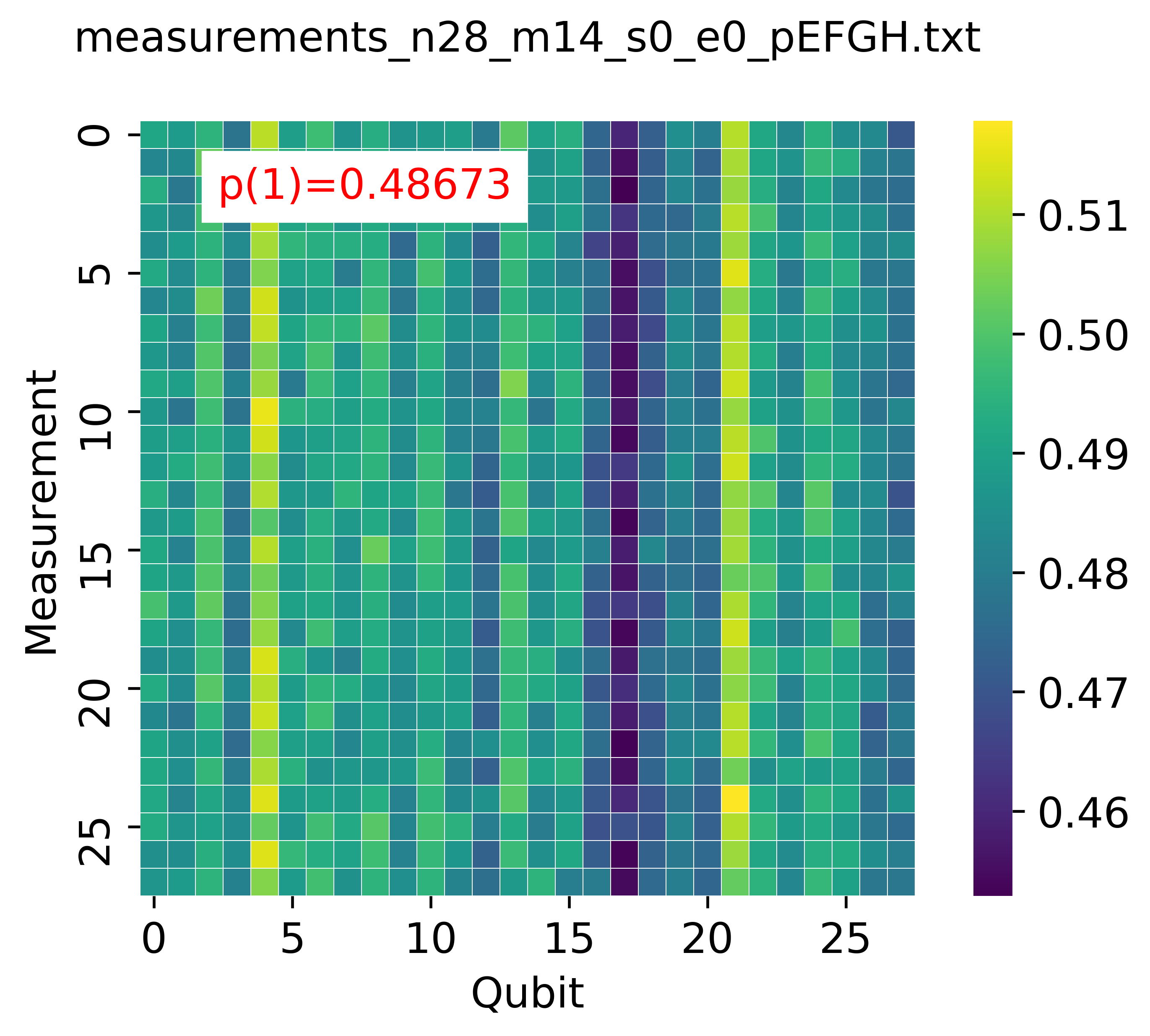}
\includegraphics[width=0.3\textwidth]{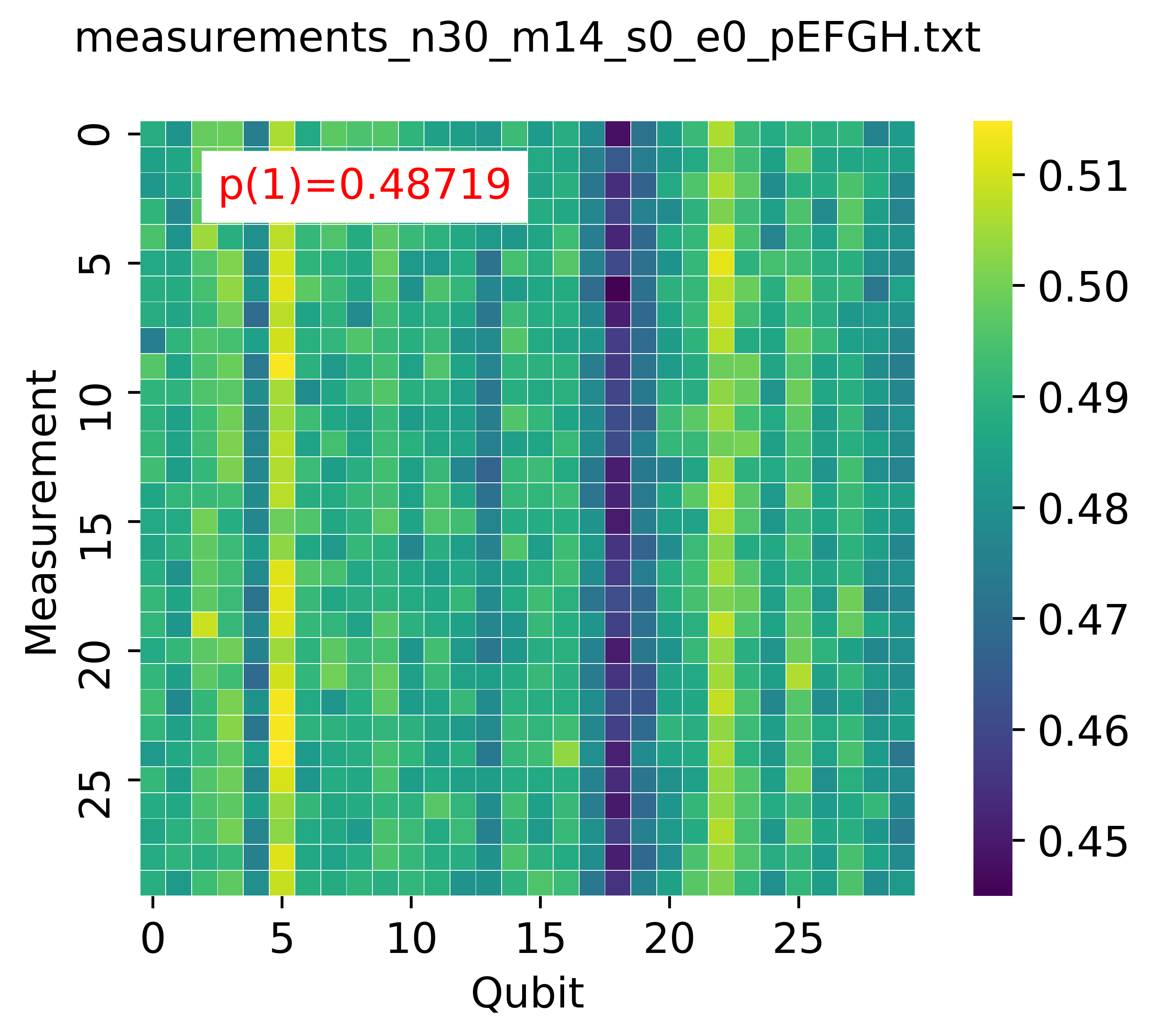}
\includegraphics[width=0.3\textwidth]{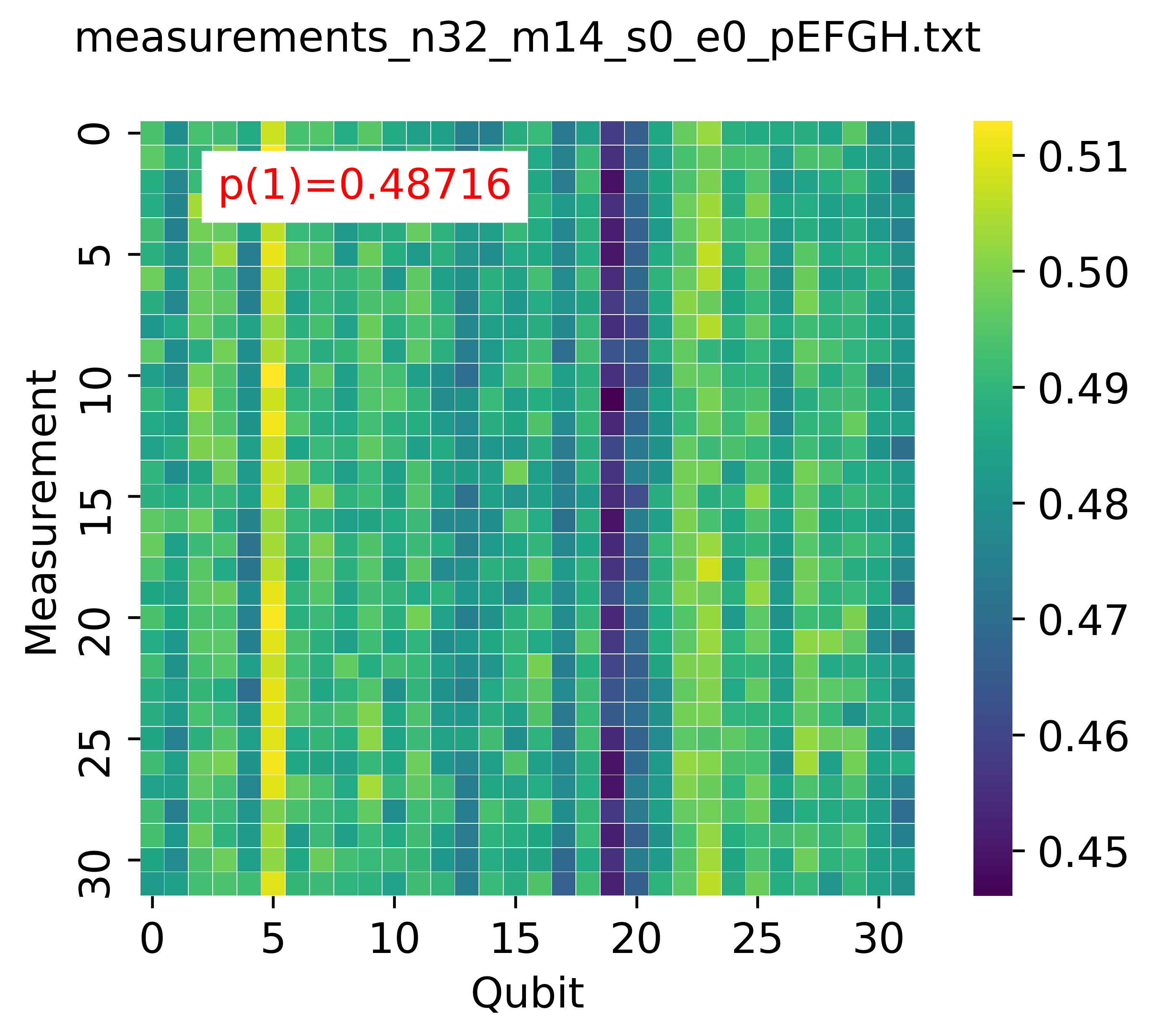}
\includegraphics[width=0.3\textwidth]{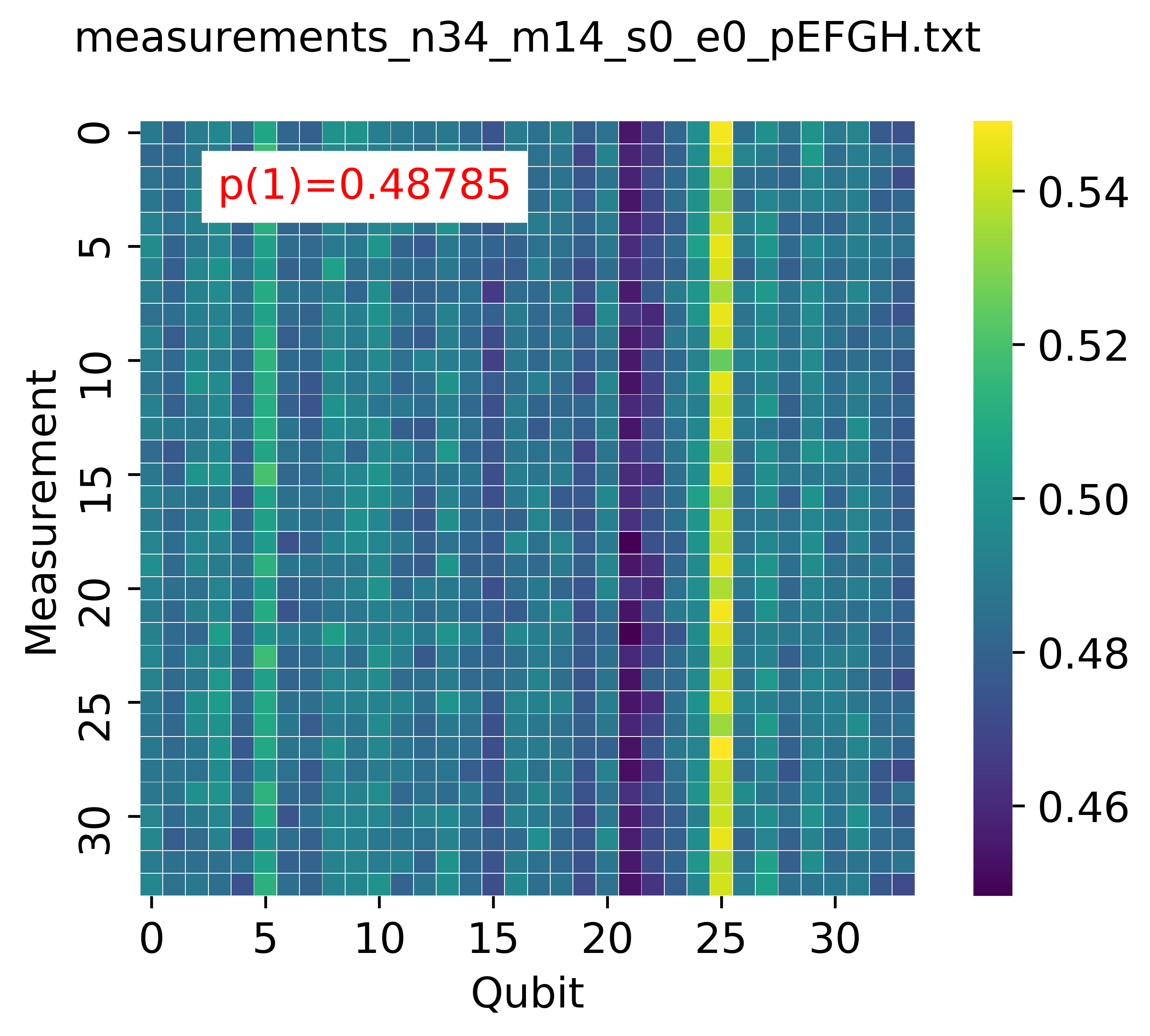}
\includegraphics[width=0.3\textwidth]{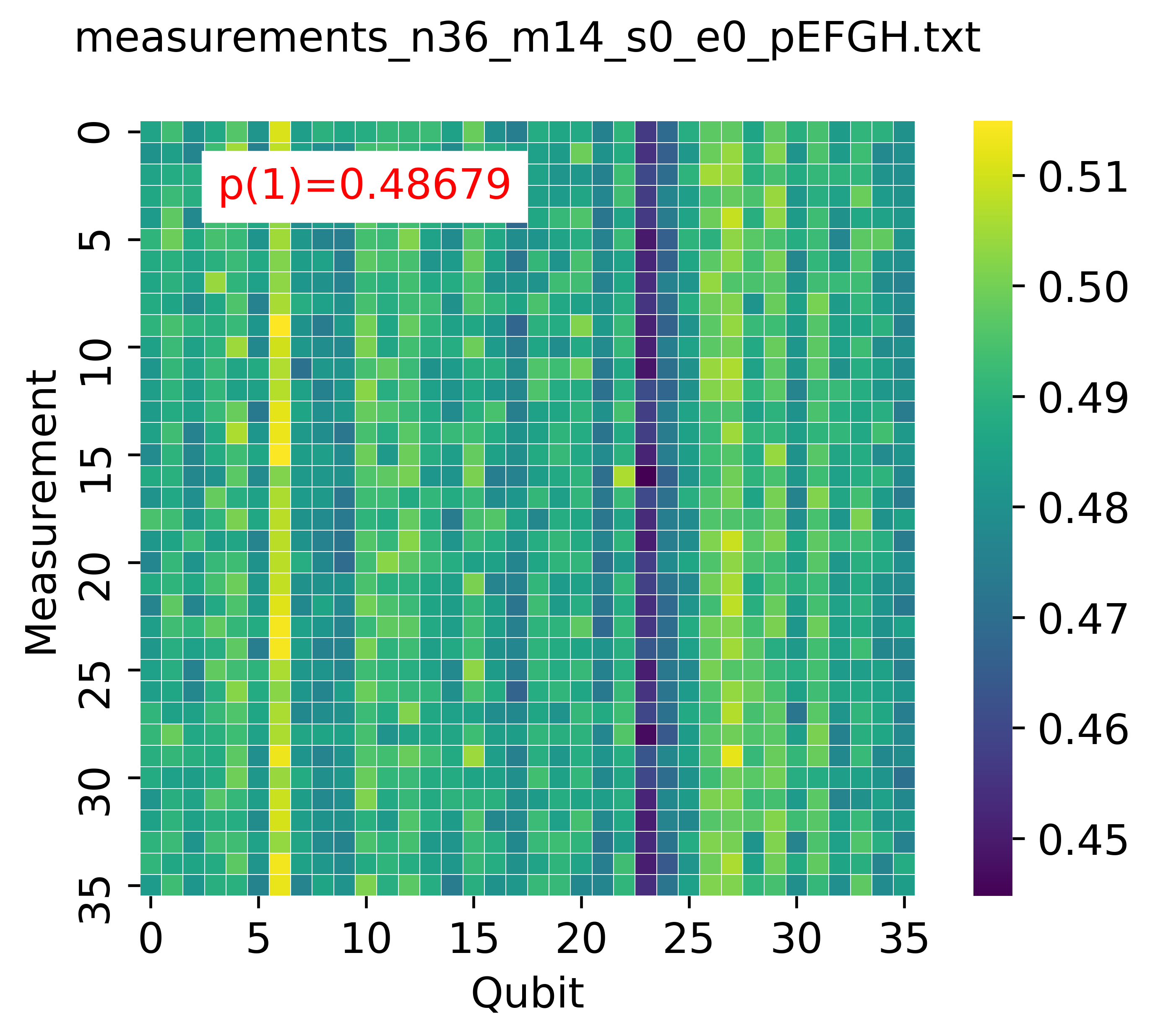}
\includegraphics[width=0.3\textwidth]{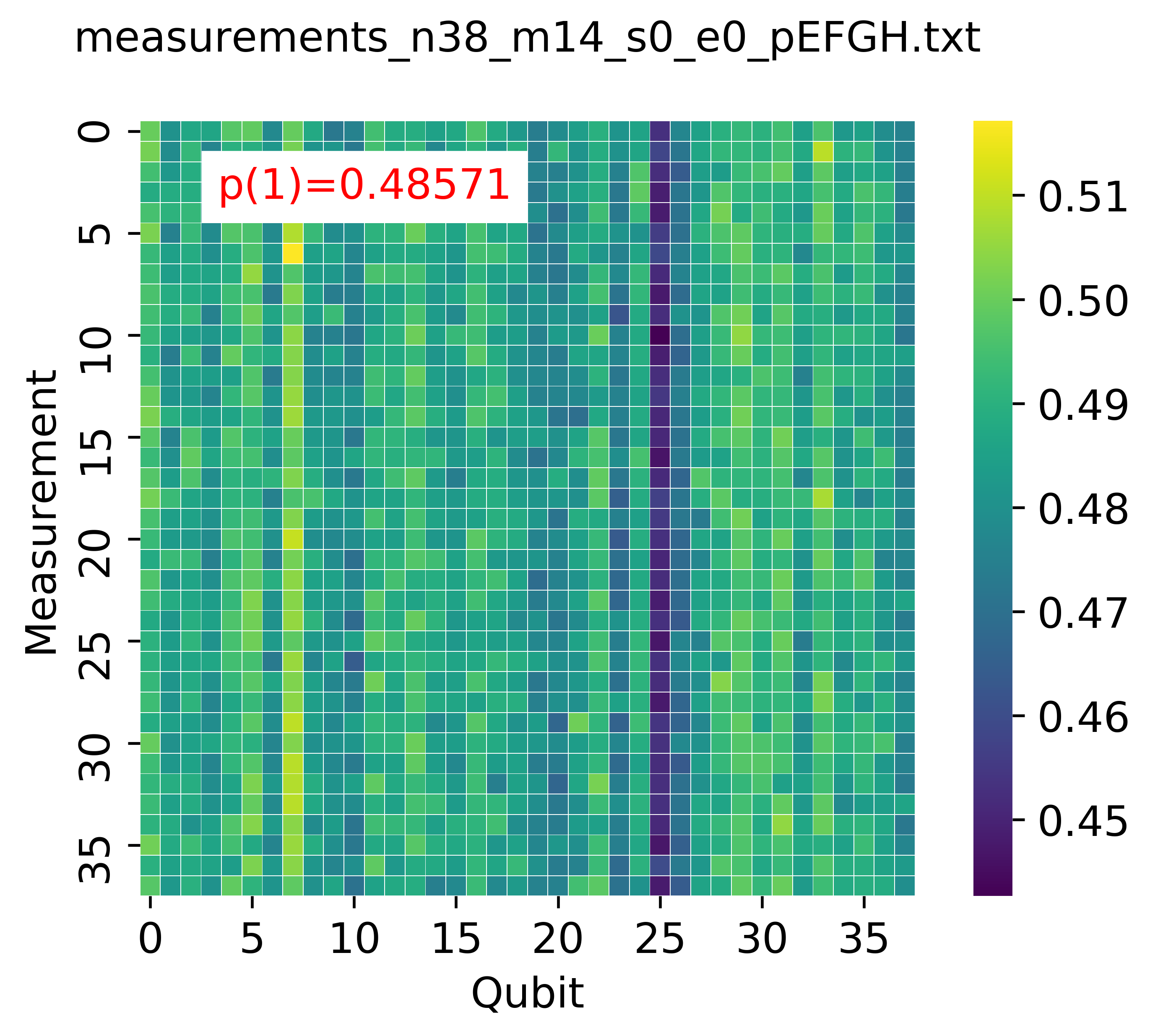}
\includegraphics[width=0.3\textwidth]{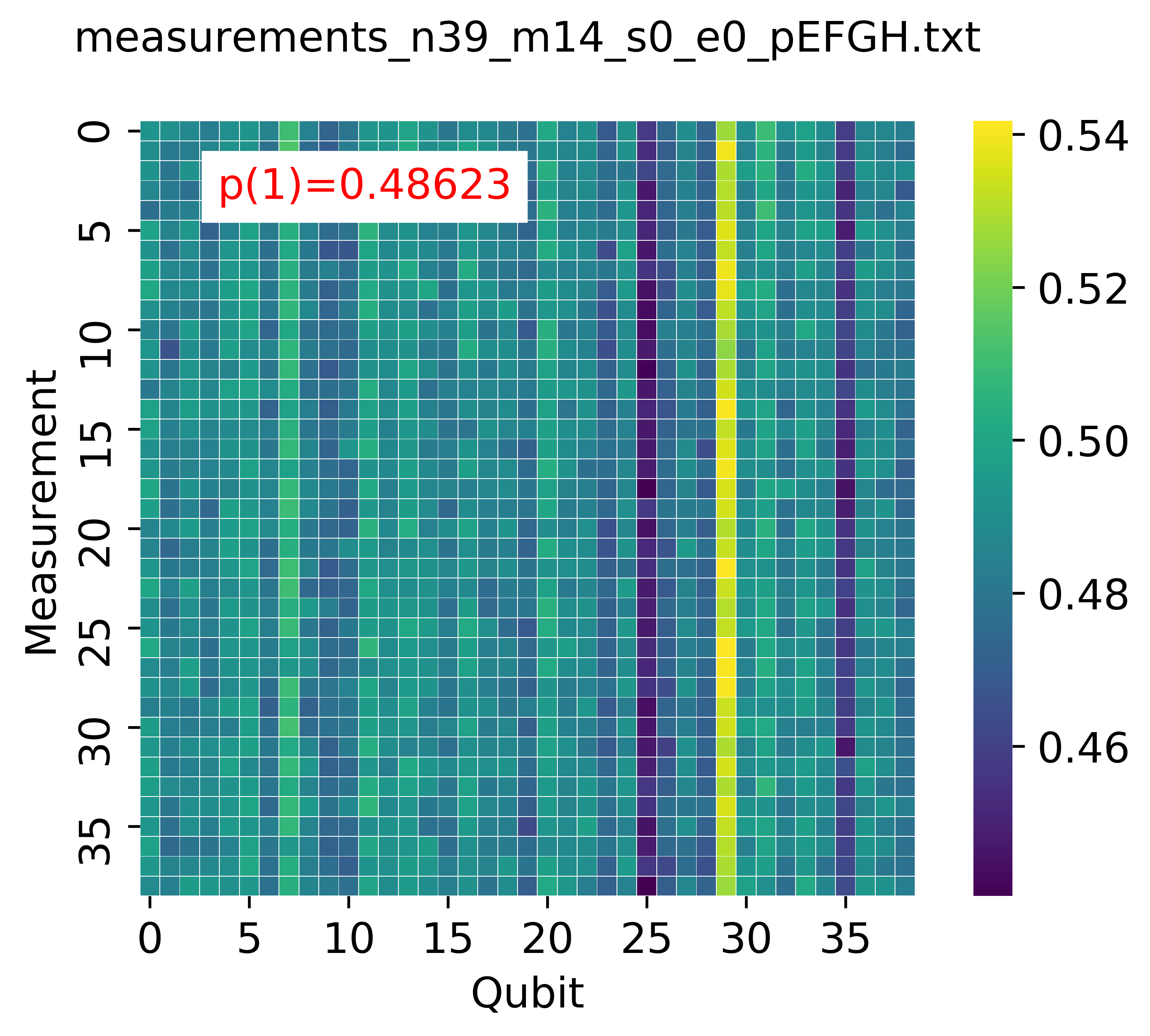}
\includegraphics[width=0.3\textwidth]{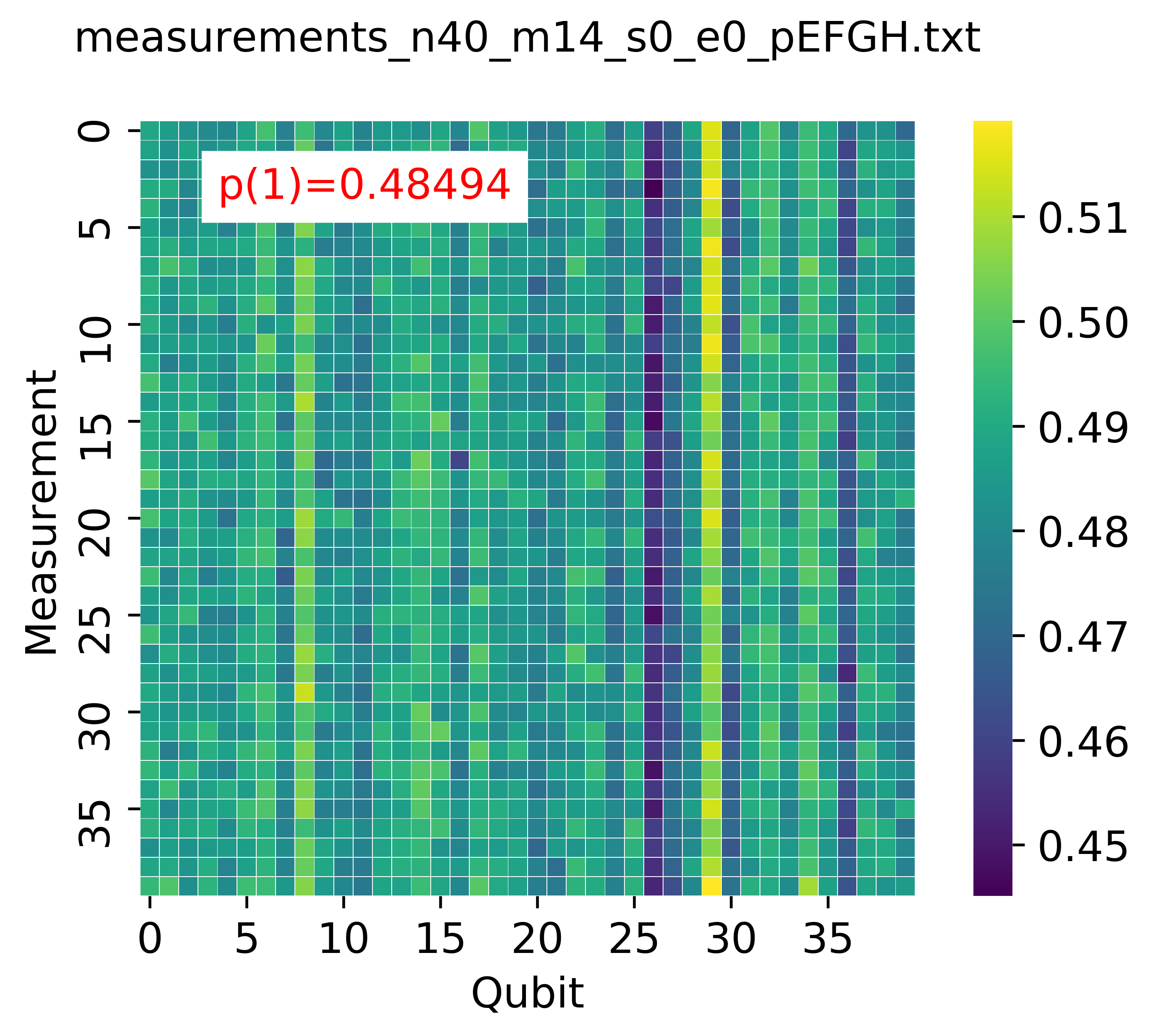}
\caption{Heat maps of Google's random bit strings for $n=14, 16, 18, 20, 22, 24, 26, 28, 30, 32, 34,
36, 38, 39, 40$ and \texttt{EFGH} activation pattern.}
\label{Heatmap_Google_EFGH1}
\end{figure*}
\vfill
\pagebreak[4]

\begin{figure*}[h]
\includegraphics[width=0.3\textwidth]{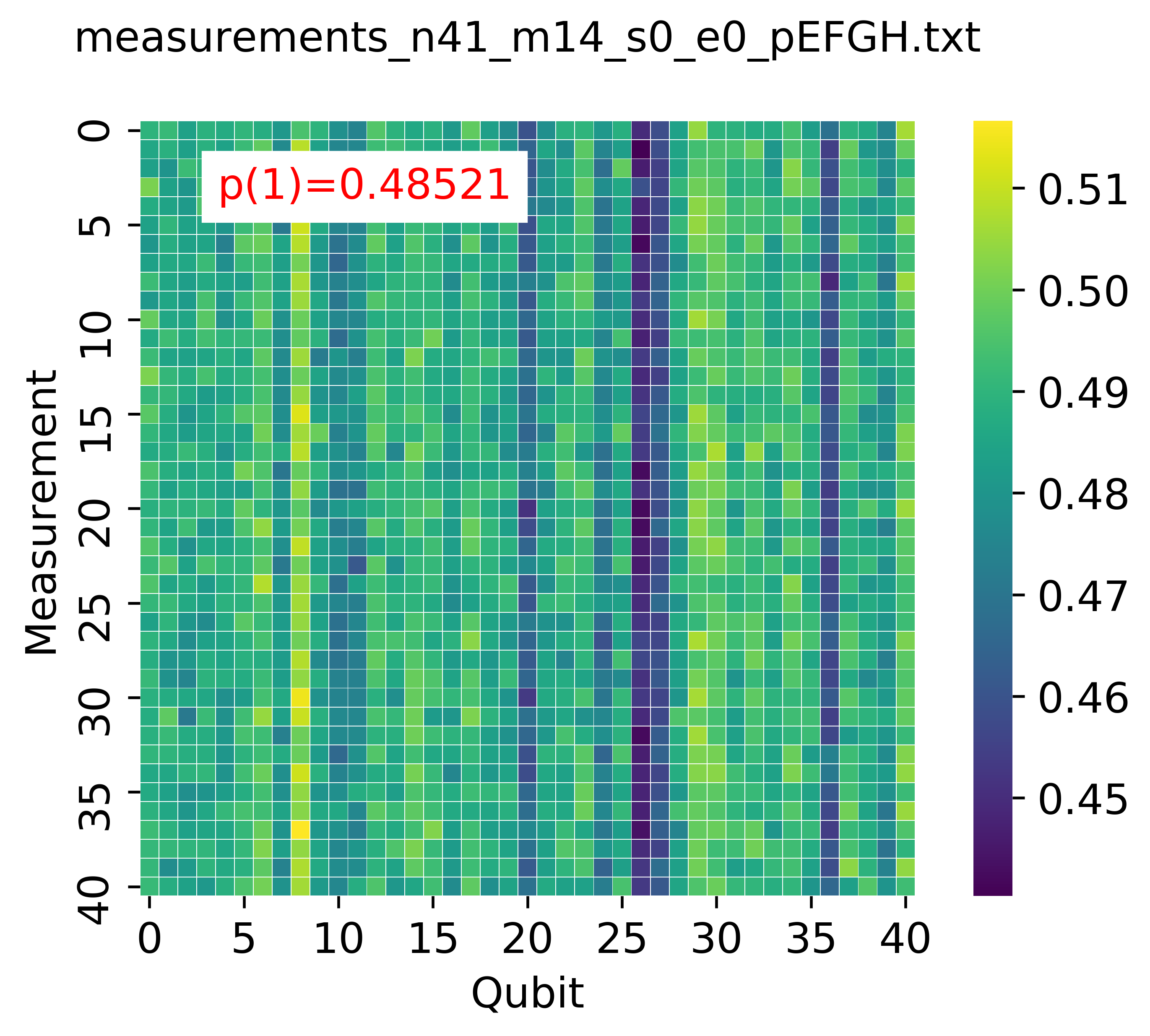}
\includegraphics[width=0.3\textwidth]{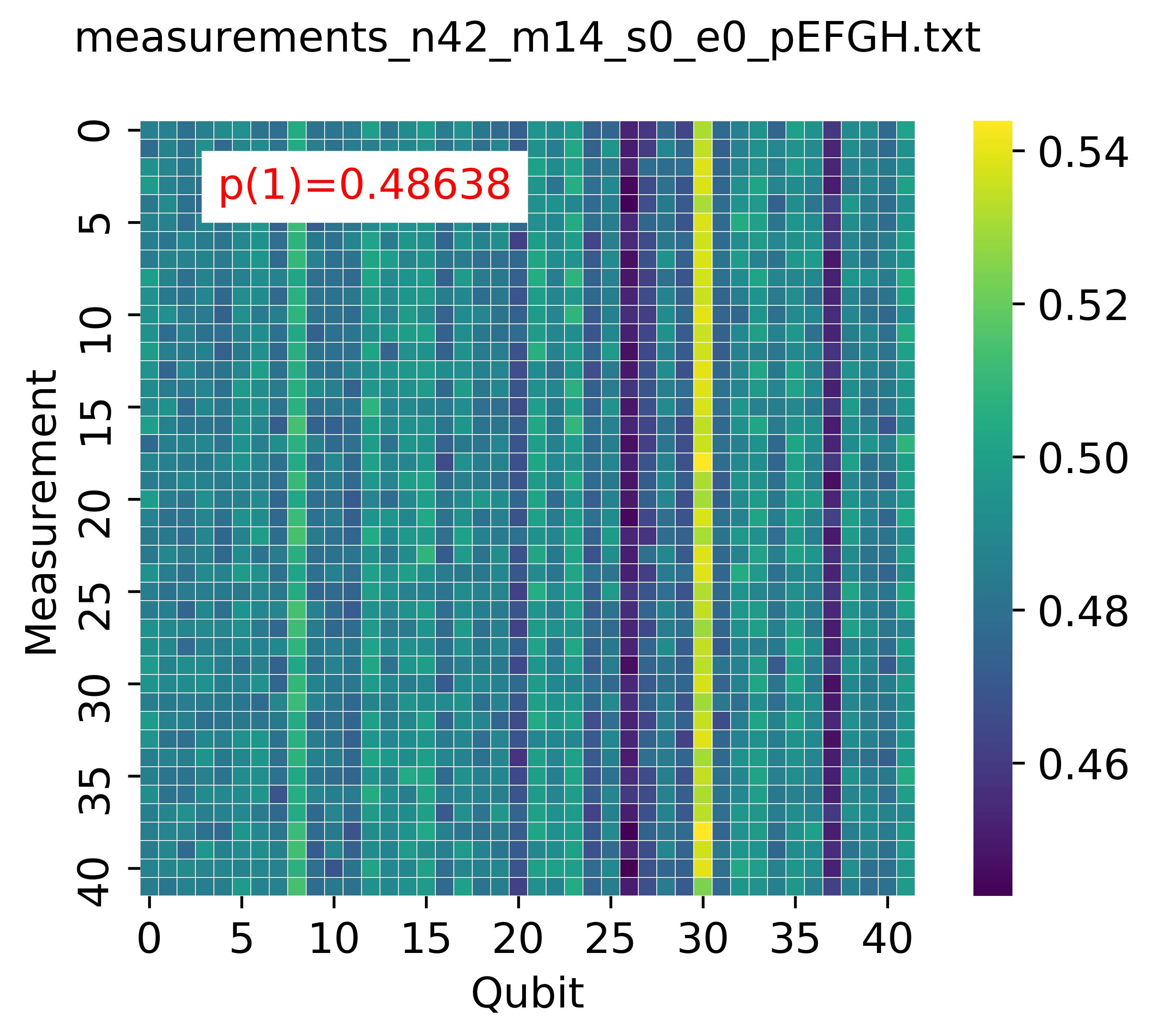}
\includegraphics[width=0.3\textwidth]{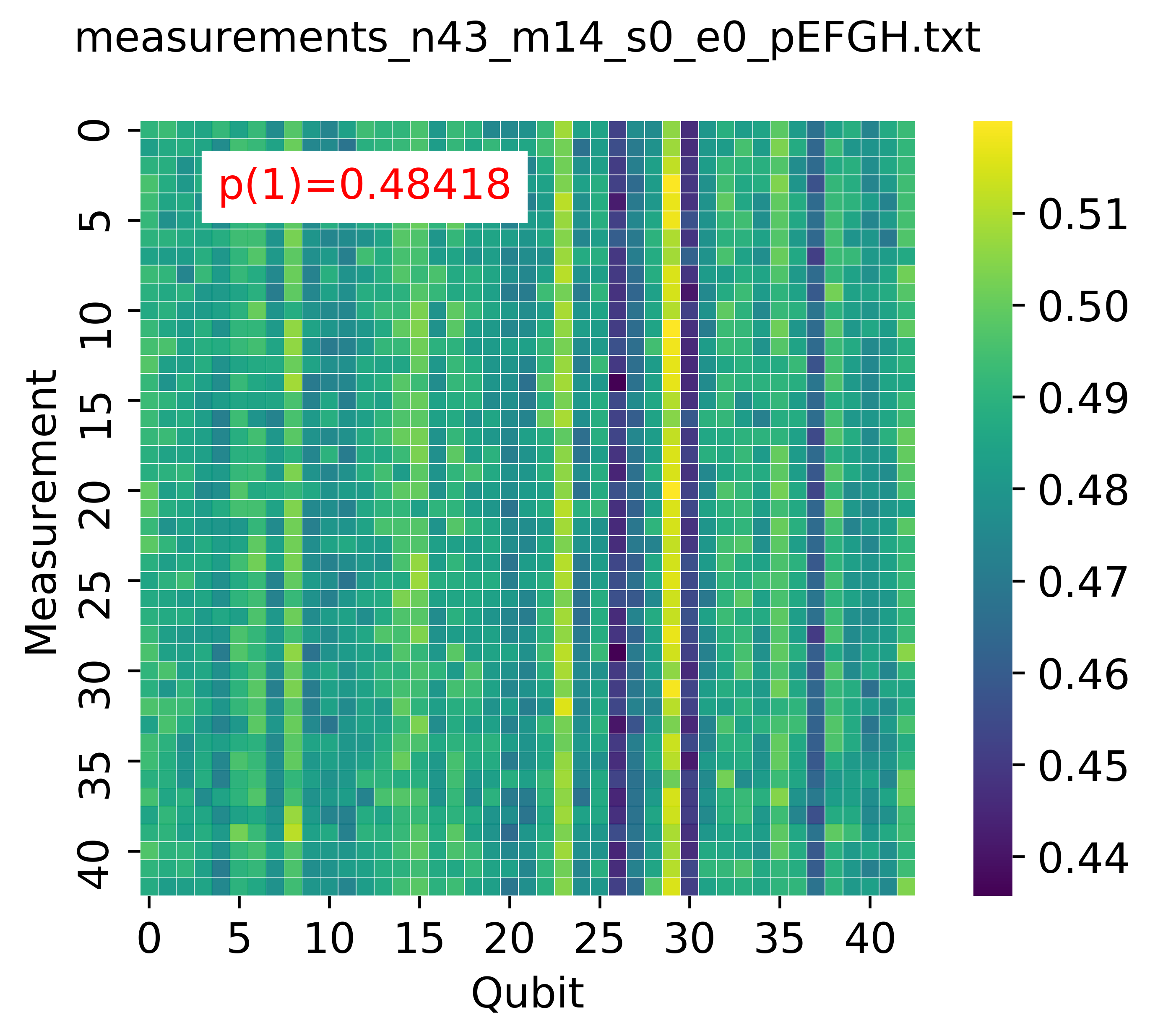}
\includegraphics[width=0.3\textwidth]{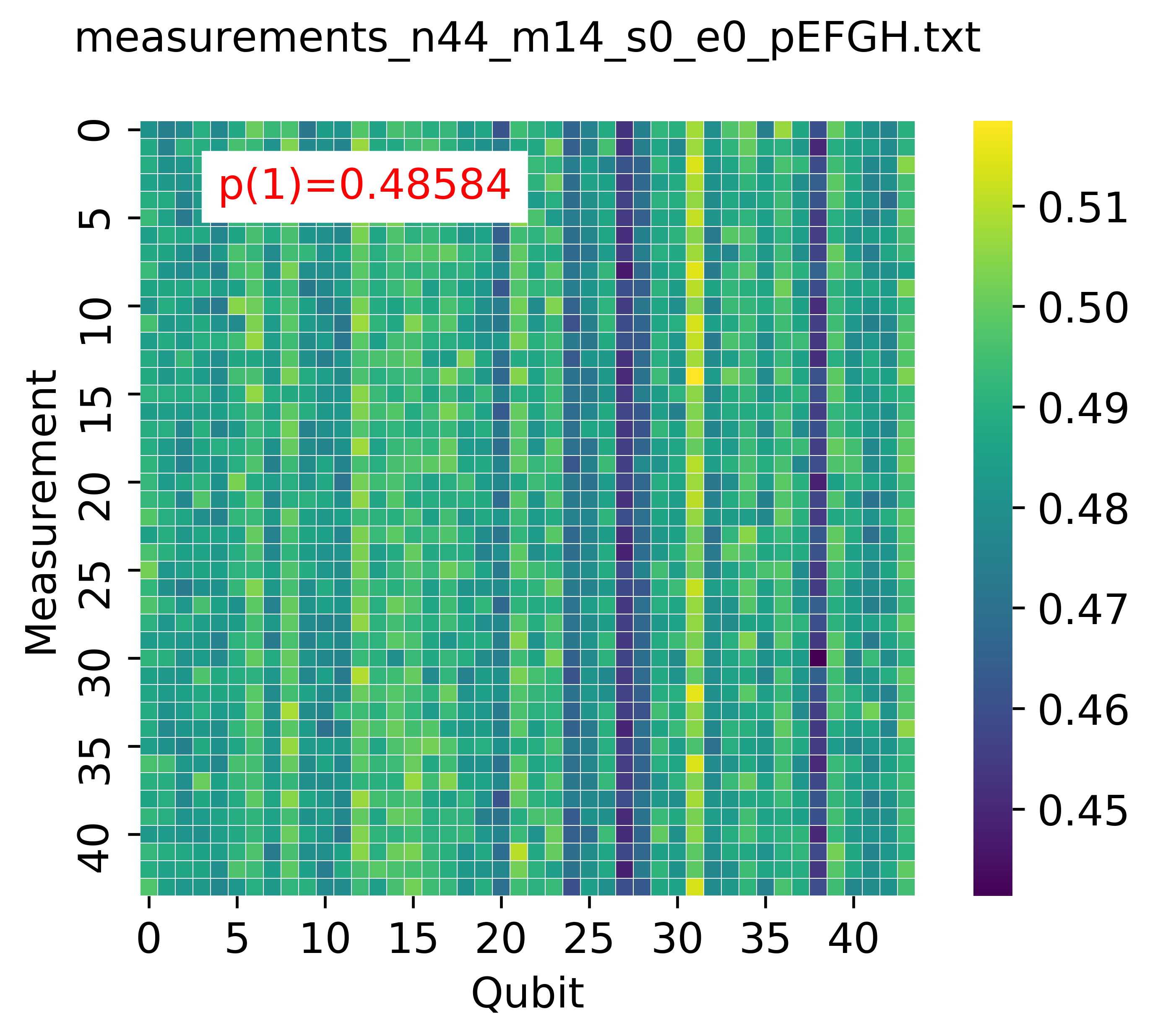}
\includegraphics[width=0.3\textwidth]{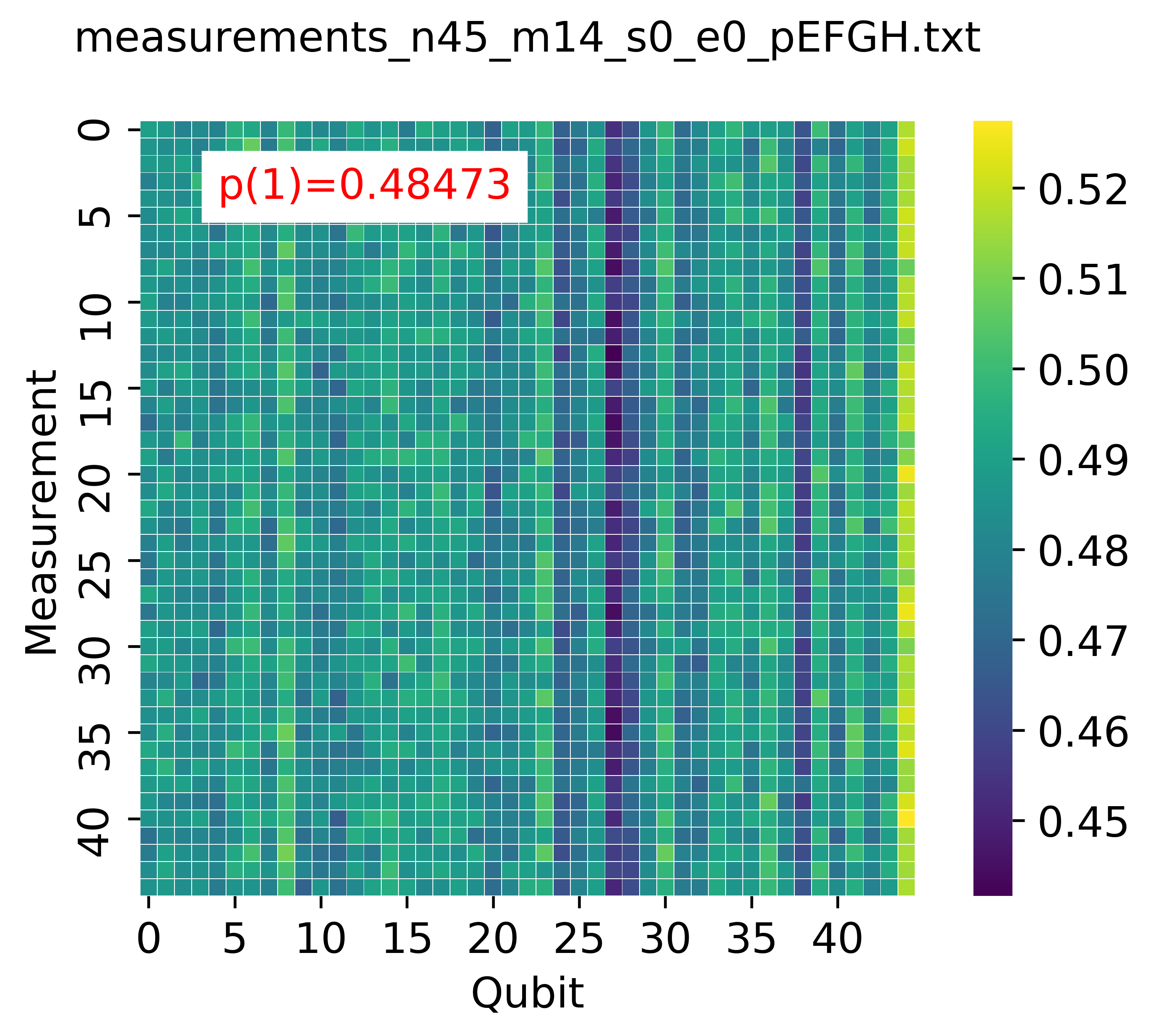}
\includegraphics[width=0.3\textwidth]{measurements_n43_m14_s0_e0_pEFGH.png}
\includegraphics[width=0.3\textwidth]{measurements_n44_m14_s0_e0_pEFGH.png}
\includegraphics[width=0.3\textwidth]{measurements_n45_m14_s0_e0_pEFGH.png}
\includegraphics[width=0.3\textwidth]{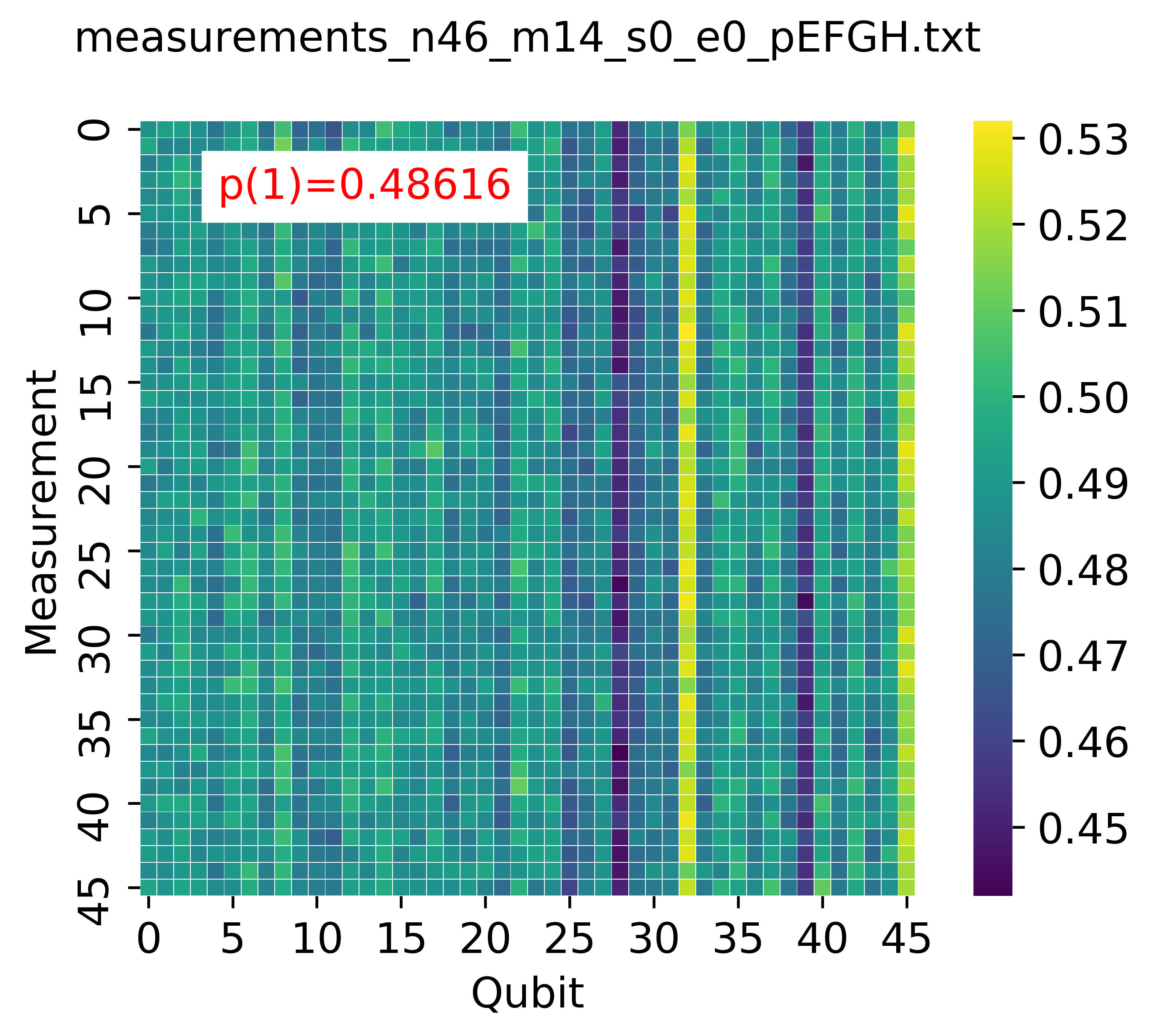}
\includegraphics[width=0.3\textwidth]{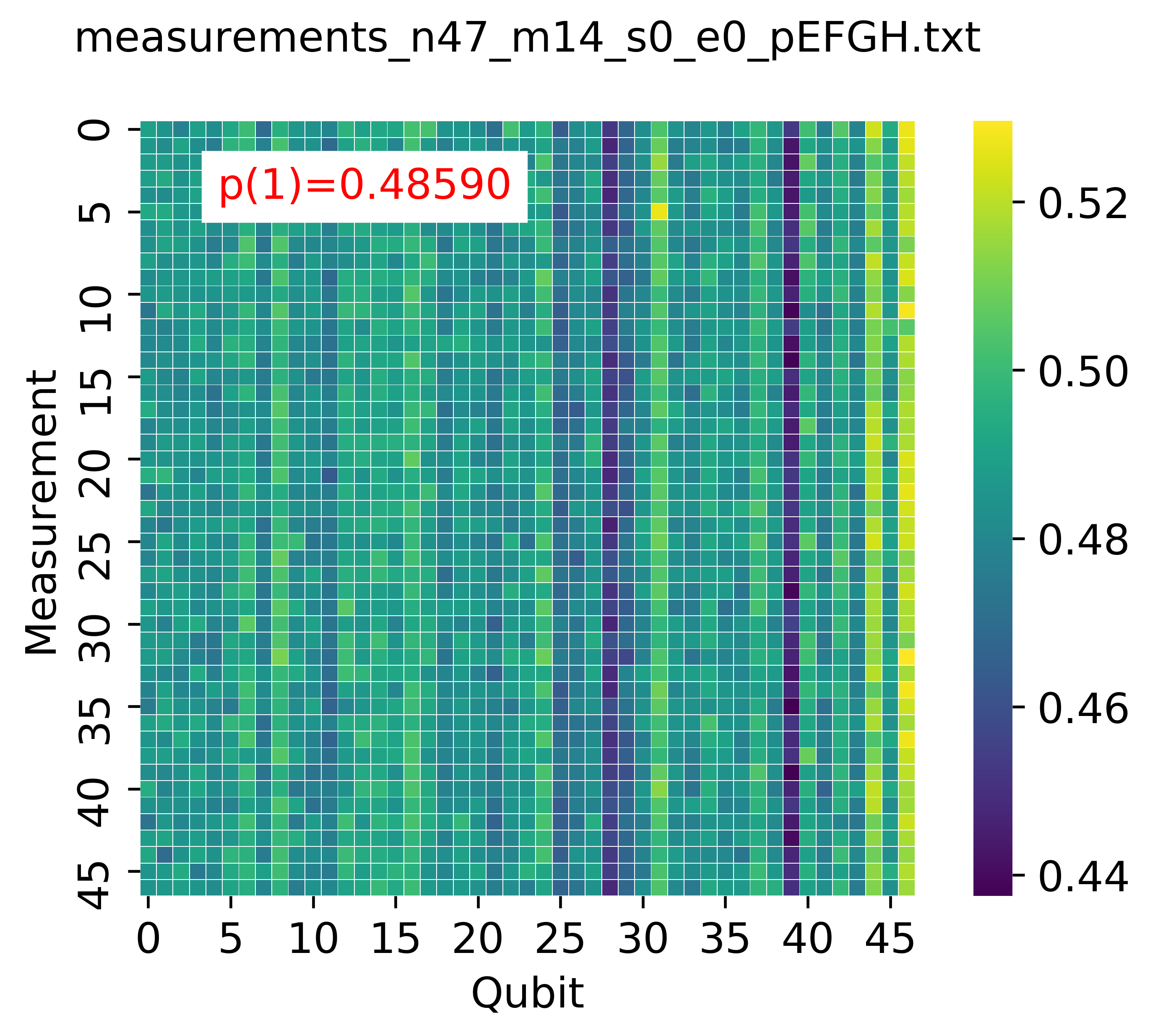}
\includegraphics[width=0.3\textwidth]{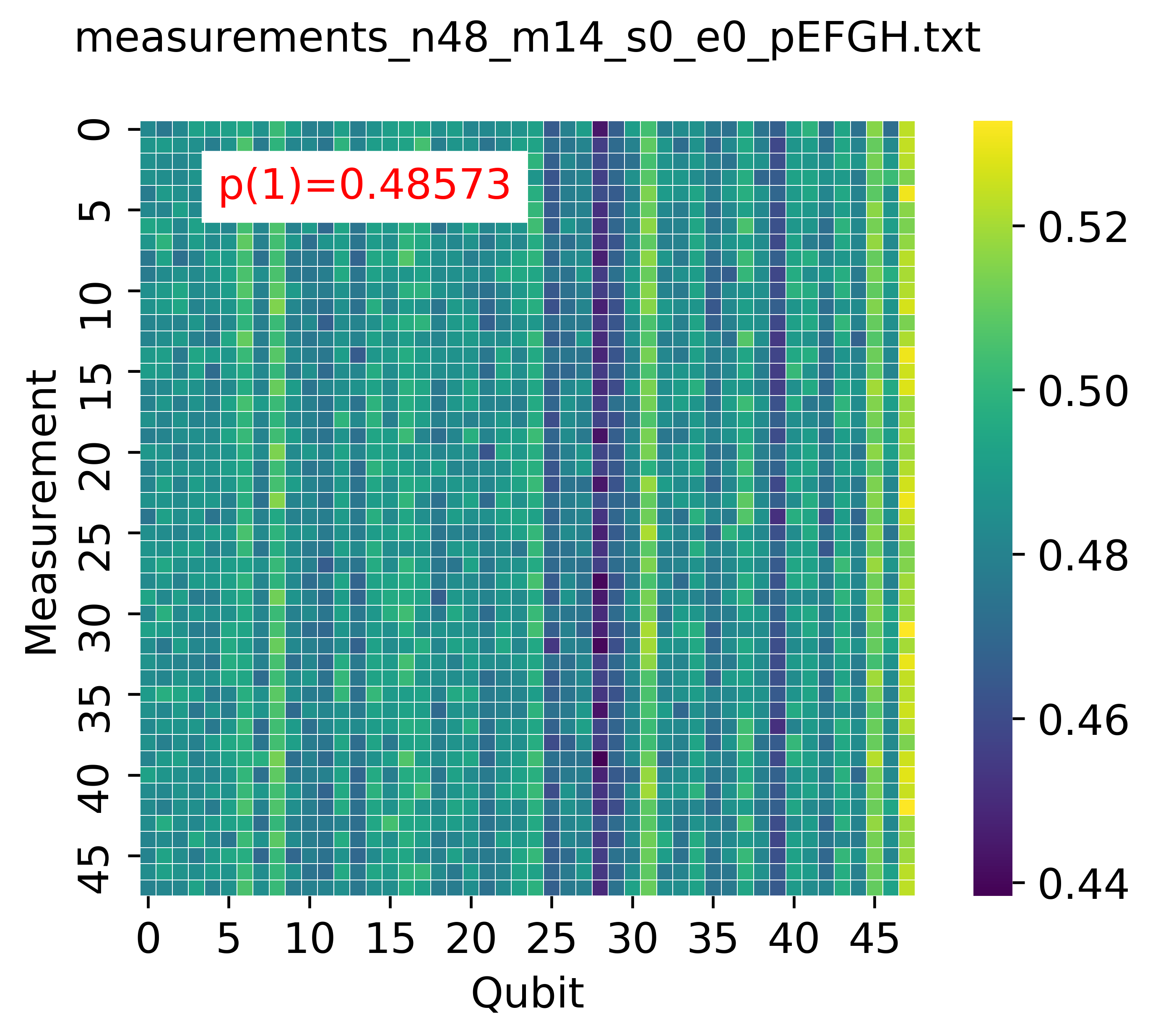}
\includegraphics[width=0.3\textwidth]{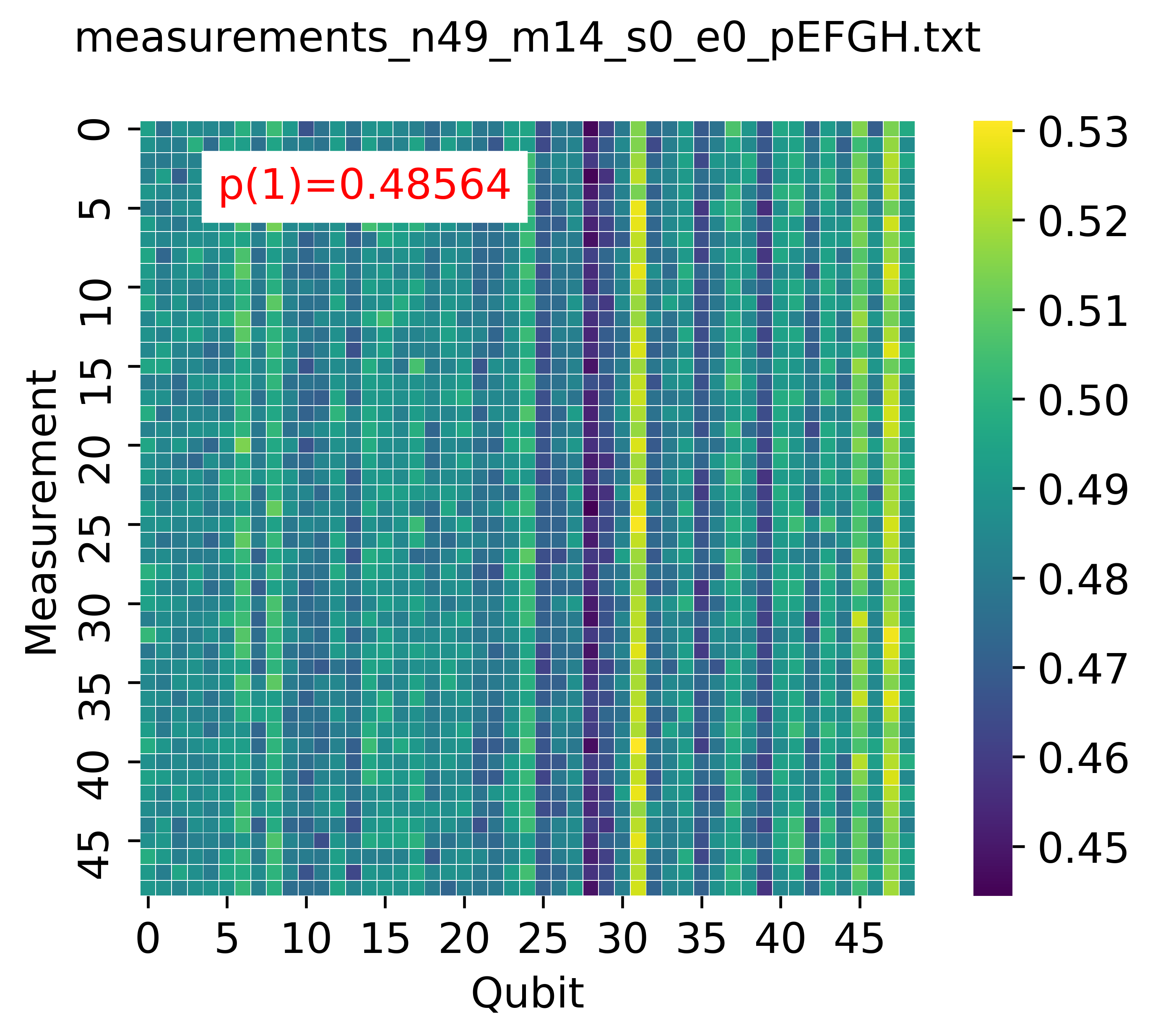}
\includegraphics[width=0.3\textwidth]{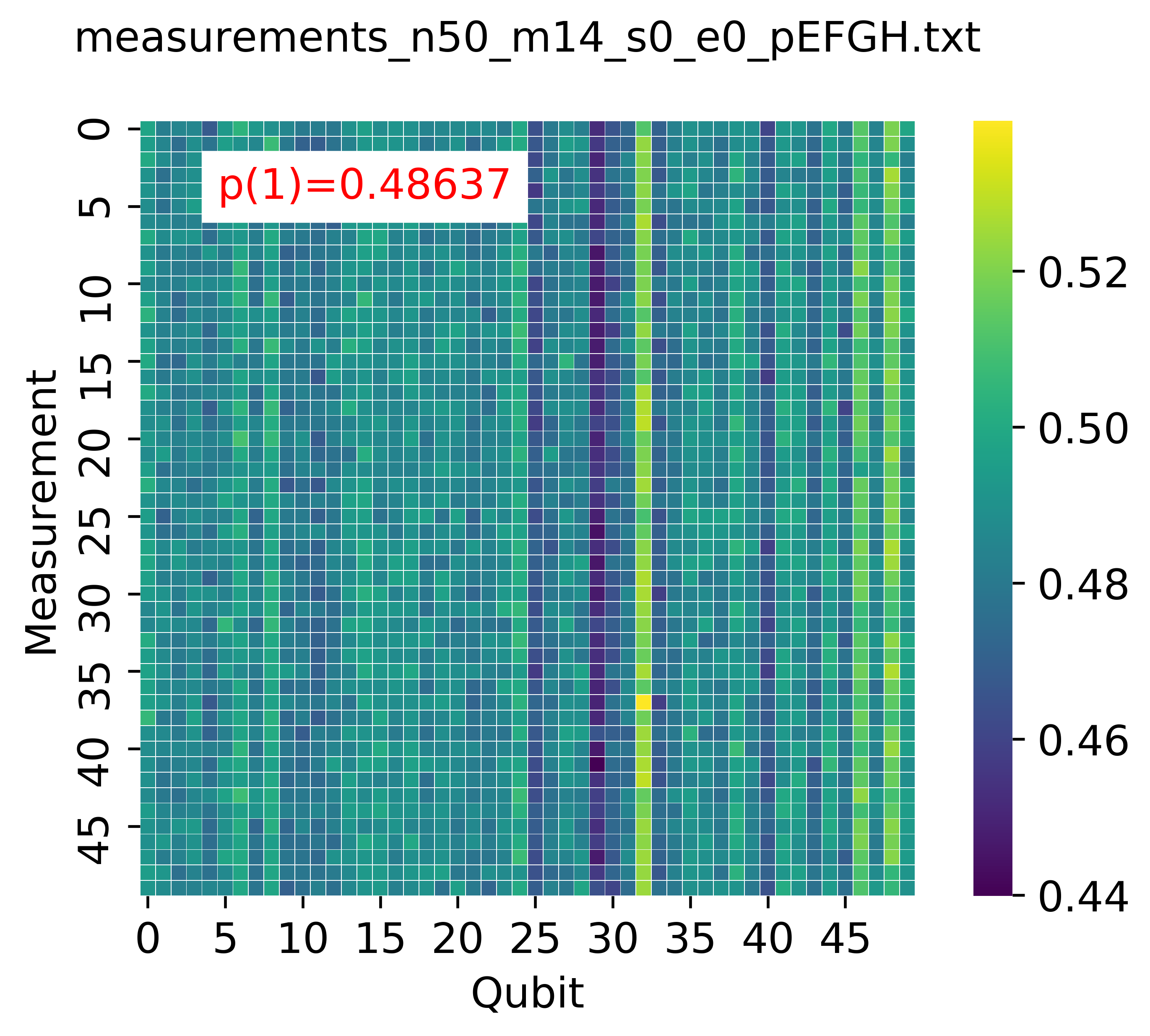}
\includegraphics[width=0.3\textwidth]{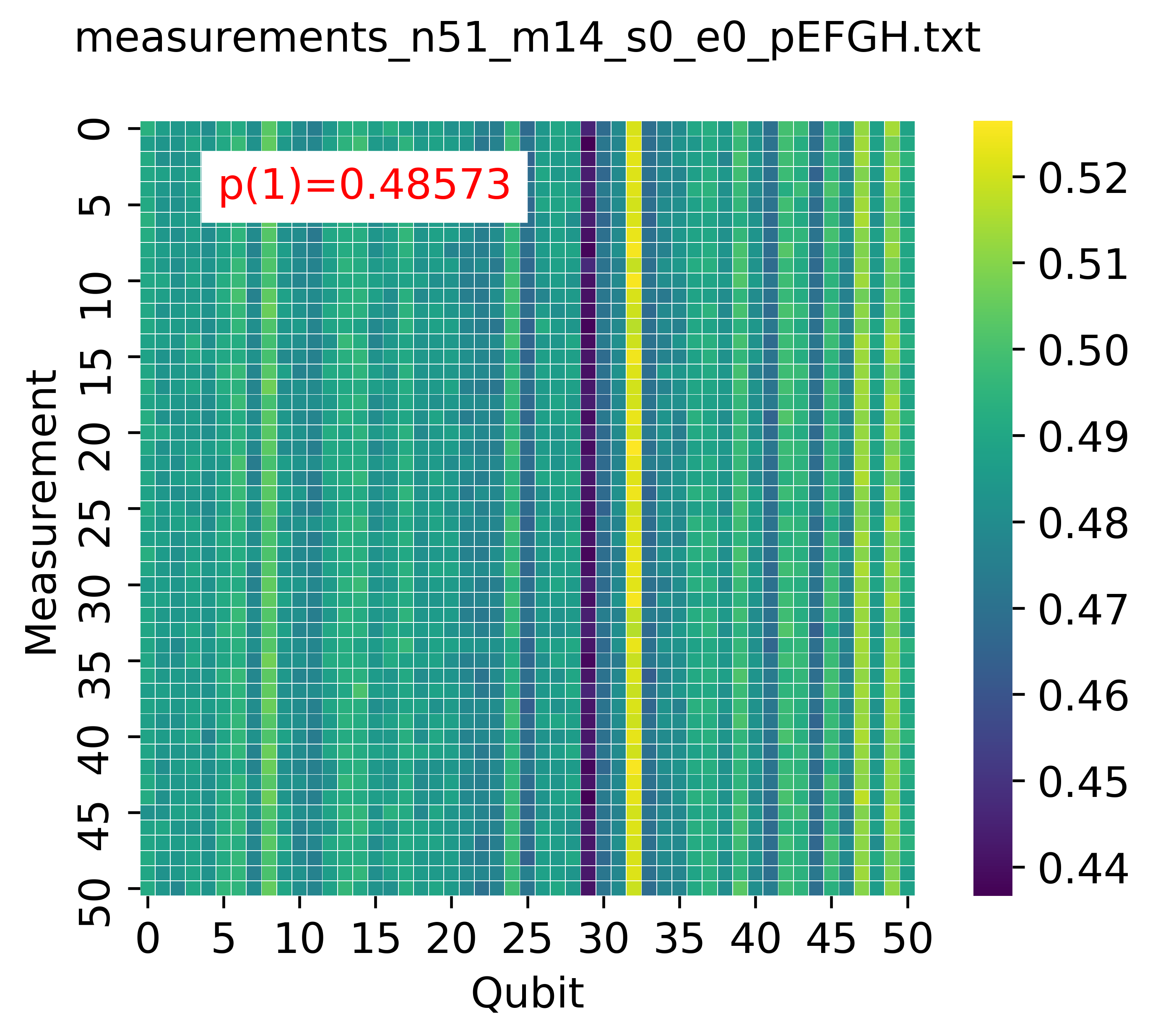}
\includegraphics[width=0.3\textwidth]{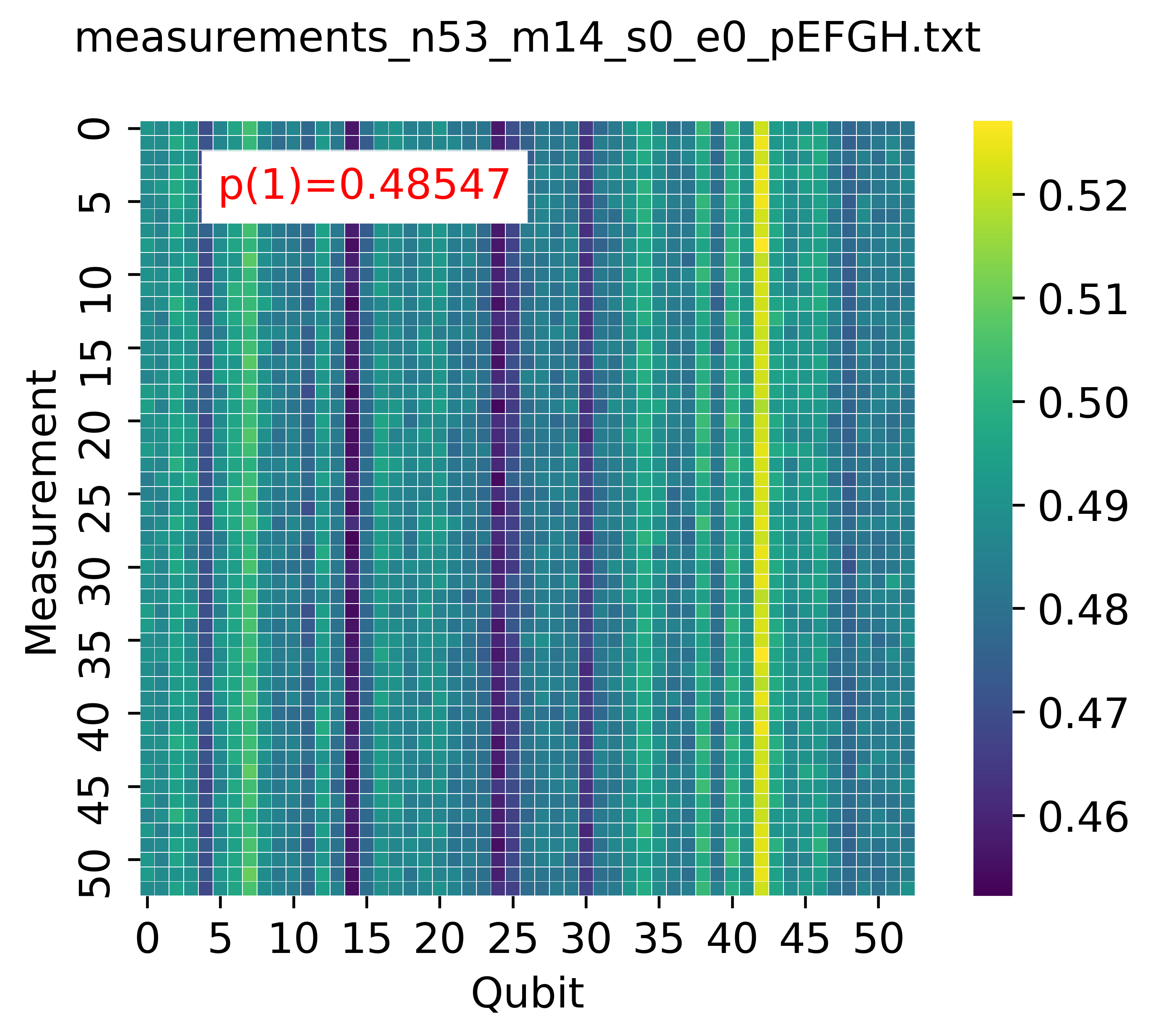}
\caption{Heat maps of Google's random bit strings for $41, 42, 43, 44, 45, 46, 47, 48, 49, 50, 51, 53$ 
and \texttt{EFGH} activation pattern.}
\label{Heatmap_Google_EFGH2}
\end{figure*}

\pagebreak[4]
\newpage

\begin{figure*}[h]
\includegraphics[width=0.3\textwidth]{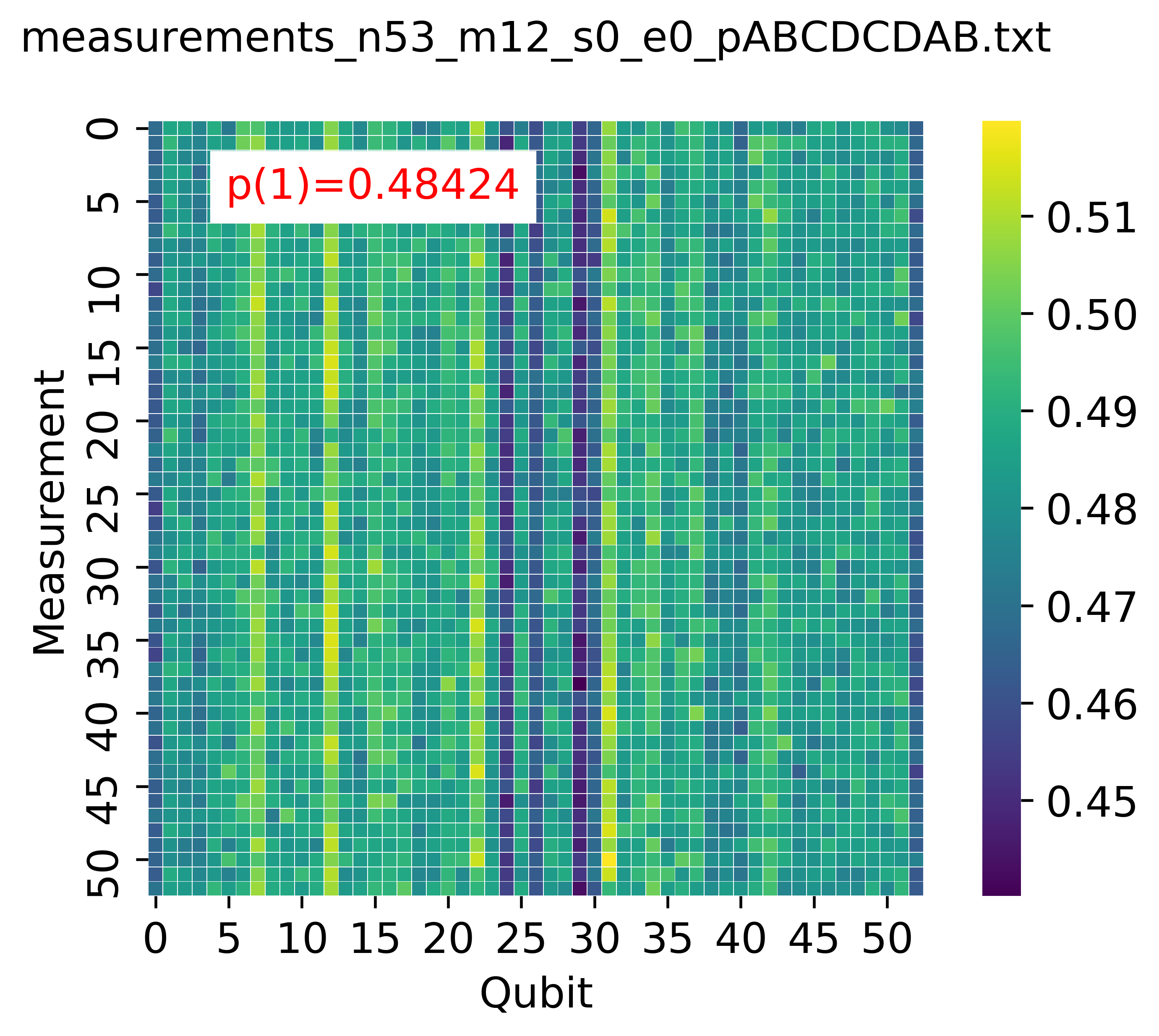}
\includegraphics[width=0.3\textwidth]{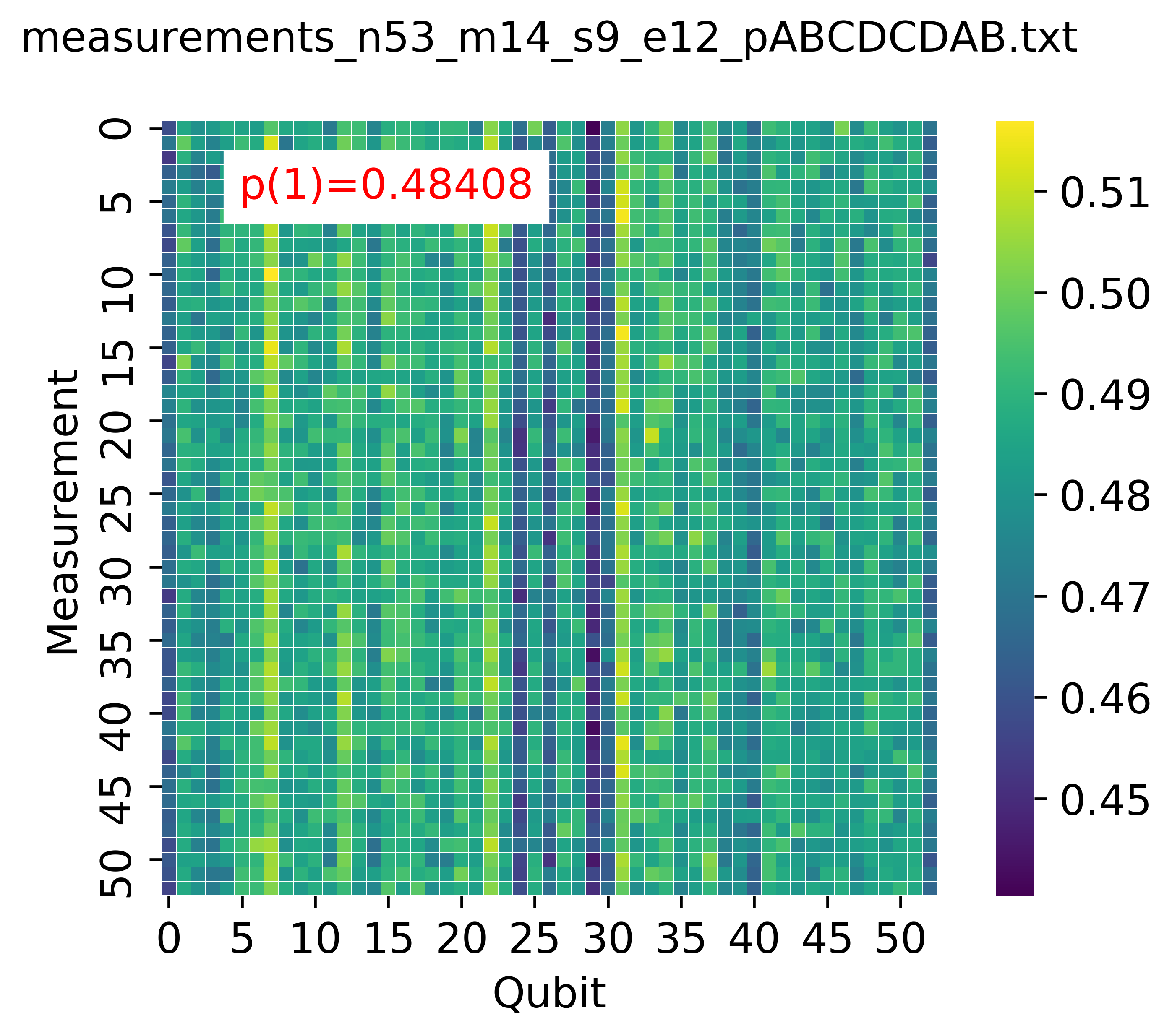}
\includegraphics[width=0.3\textwidth]{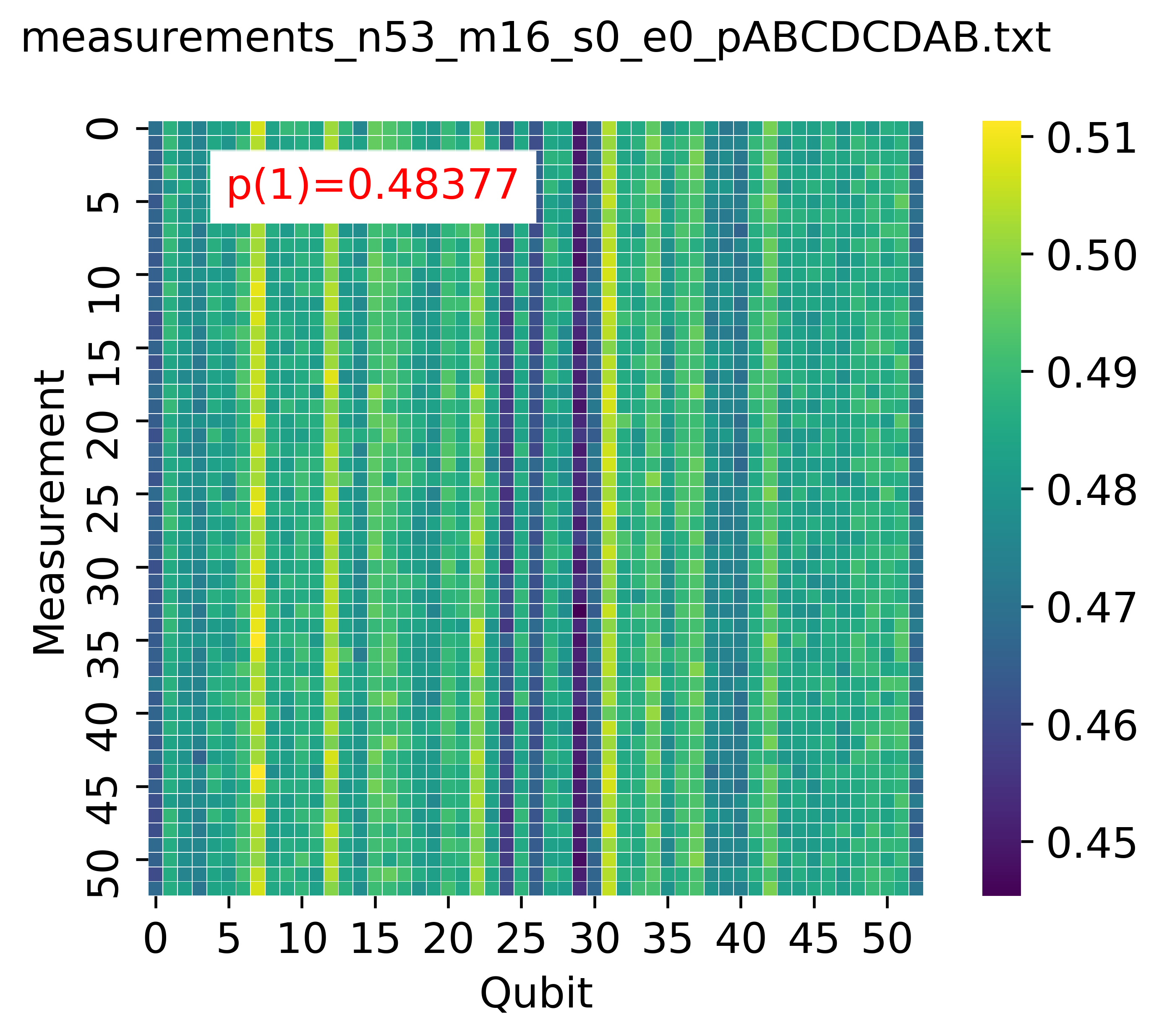}
\includegraphics[width=0.3\textwidth]{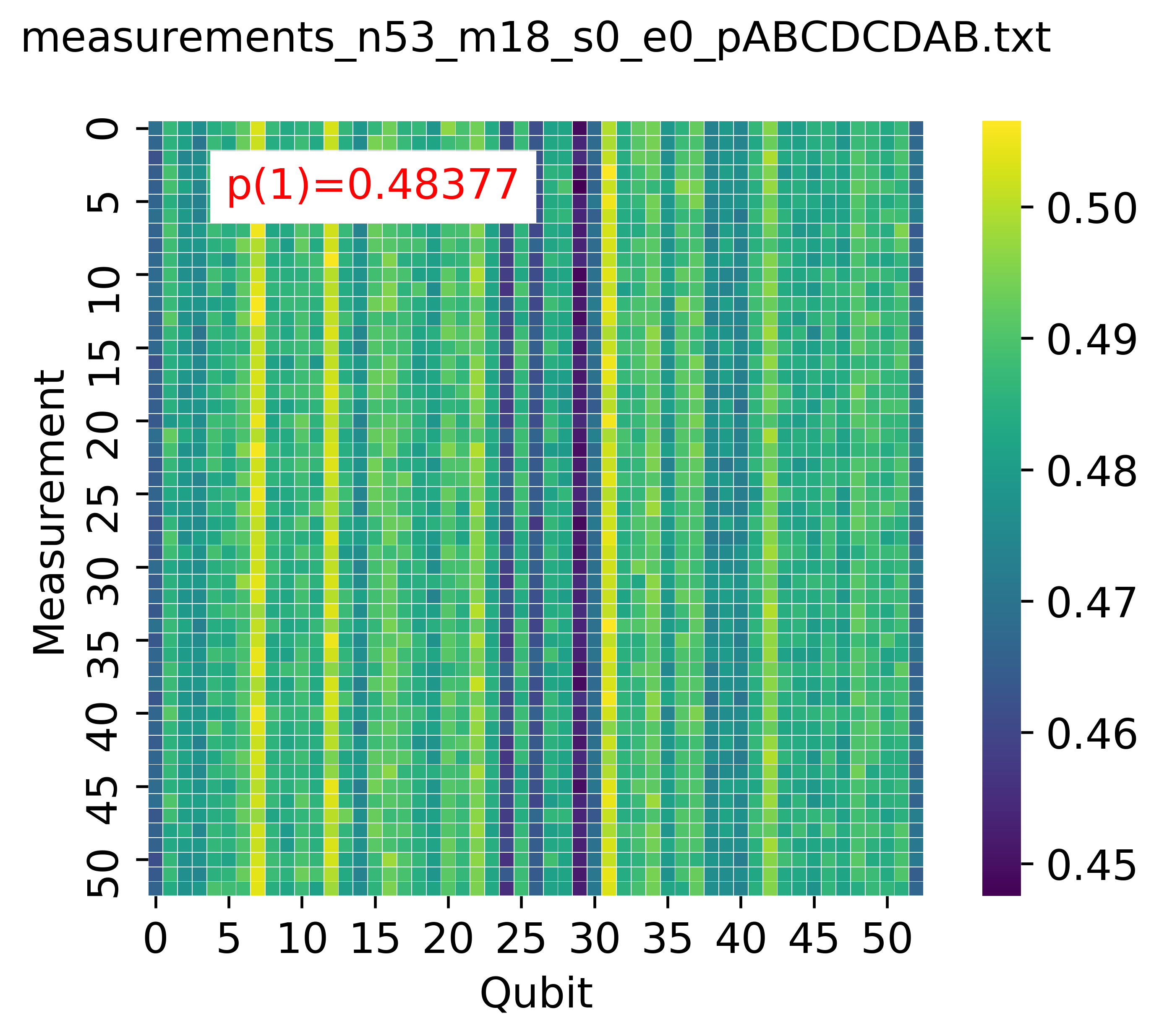}
\includegraphics[width=0.3\textwidth]{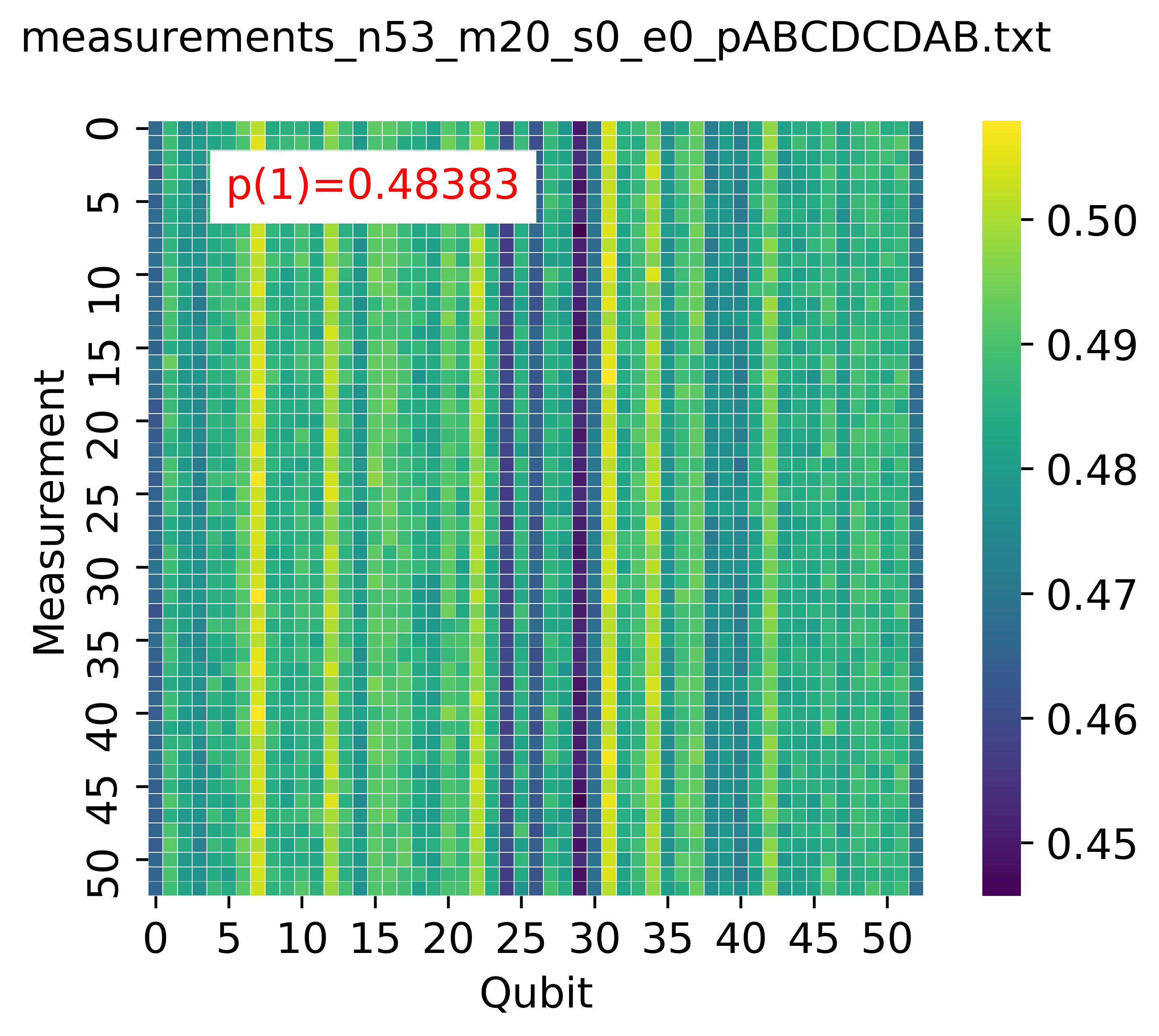}
\includegraphics[width=0.3\textwidth]{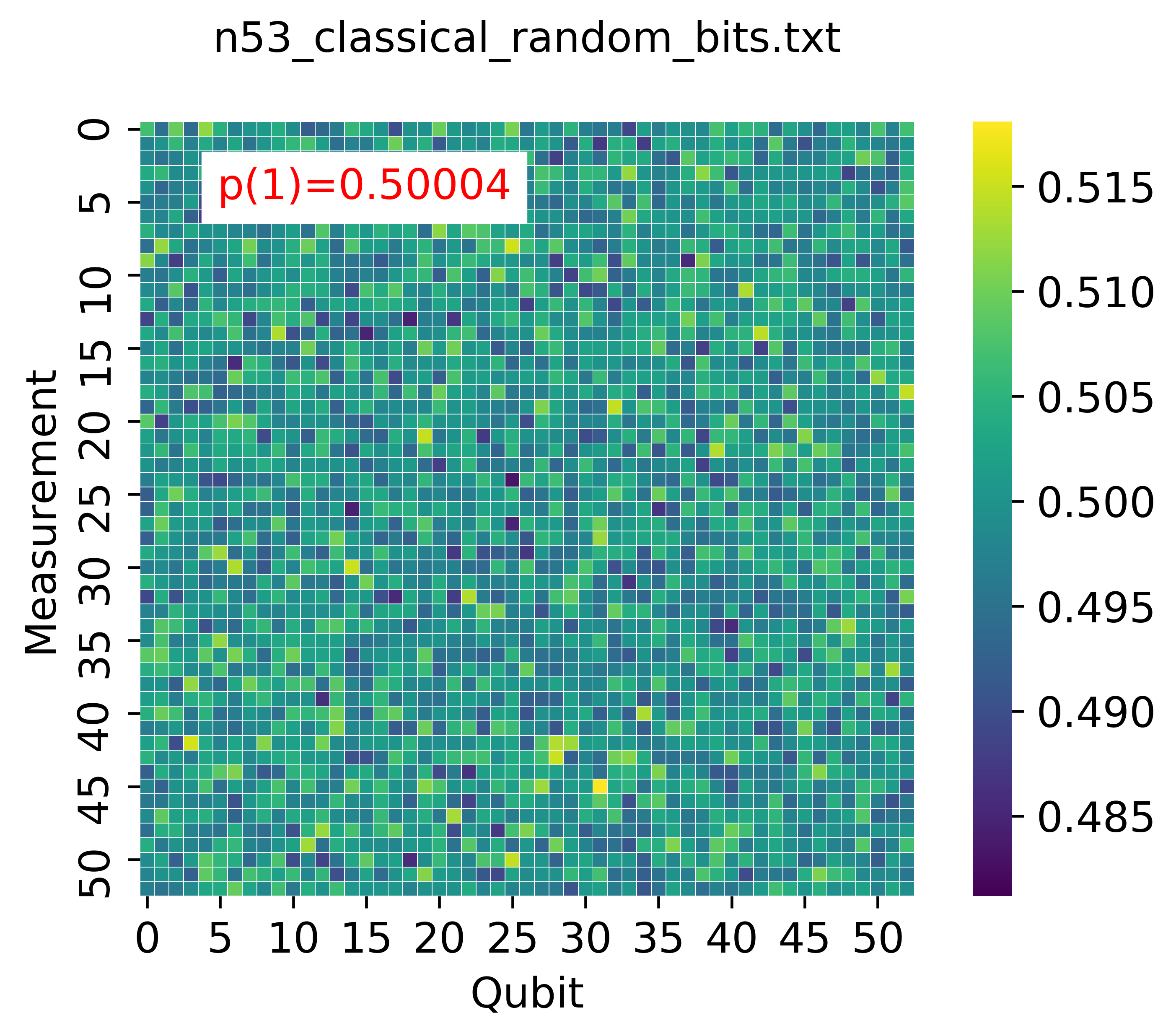}
\caption{Heat maps of Google's random bit strings for $n=53$, \texttt{ABCDCDAB} activation pattern, 
and cycle $m=12,14,16,18,20$.}
\label{Heatmap_Google_n53_ABCDCDAB}
\end{figure*}

\pagebreak
\clearpage

\subsection{NIST Statistical Test for Random Numbers} 
The randomness of Google's random bit-strings are tested using the NIST statistical test suite for random and pseudo-random number
generators for cryptographic applications~\cite{NIST2010}. Instead of the NIST C code, {\tt NIST SP 800-22}, 
the Python implementation of the NIST statistical test suite developed by Ang~\cite{Ang2019} is used because of the easy graphic 
user-interface, as shown in Fig.~\ref{python_NIST}. Table~\ref{Random_number_test} shows the results of the NIST random number
tests for classical random bits for $n=53$, Qiskit random circuits for $n=12$, the Haar measure sample for $n=12$, 
and Google's data from $n=12$ to $n=53$. Most Google's data do not pass the frequency test, run test, and overlapping template
matching, approximate entropy test, and cumulative sum tests. We performed the NIST statistical test for all Google data and
the result files \verb|Random_test.zip| are attached.

\begin{figure}[h]
\includegraphics[width=0.8\textwidth]{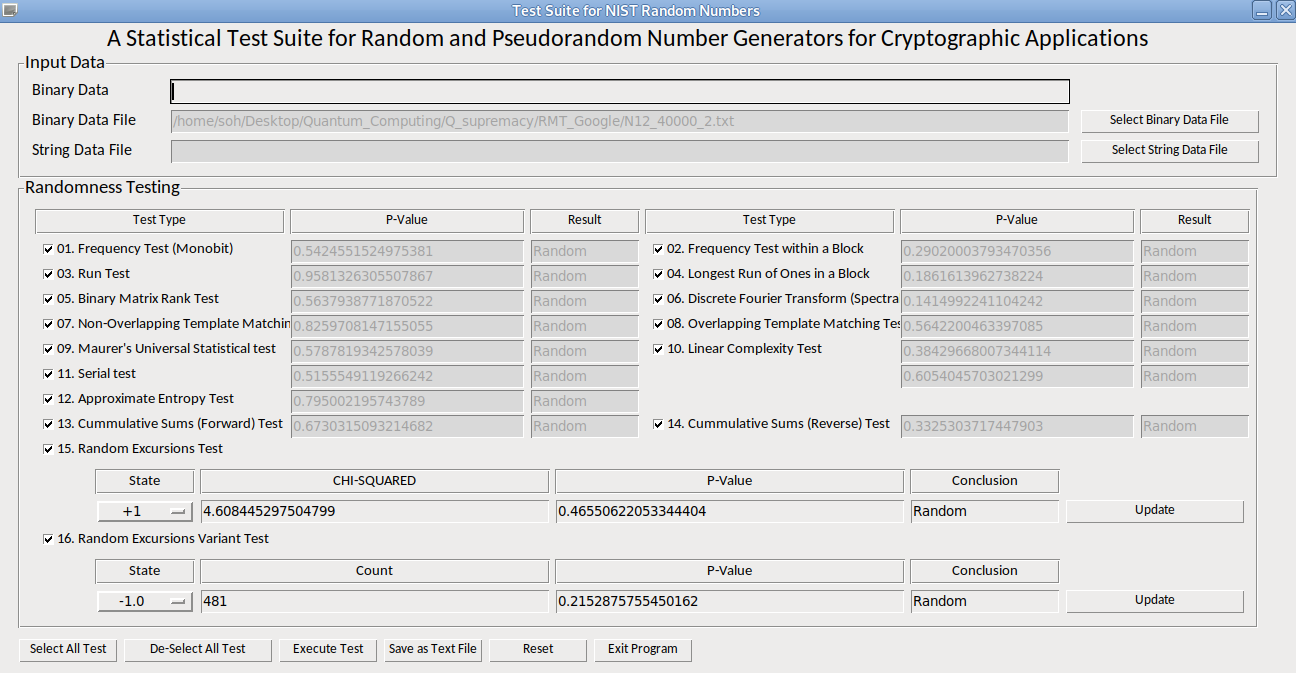}
\caption{The screen shot of a python implementation of the NIST statistical test suite~\cite{Ang2019}.}
\label{python_NIST}
\end{figure}
\vfill\pagebreak

\begin{table*}[t]
\begin{tabular}{l 
       p{14pt} p{14pt} p{14pt} p{14pt} p{14pt} p{14pt} p{14pt} p{14pt} p{14pt} p{14pt} 
       p{14pt} p{14pt} p{14pt} p{14pt} p{14pt} p{14pt} p{14pt} p{14pt}} 
       \hline\hline\\[-5pt]
     &\phantom{~} 
     &\rot[90]{01. Frequency test (mono-bit)}          
     &\rot[90]{02. Frequency test within a block}
     &\rot[90]{03. Run test}
     &\rot[90]{04. Longest run of ones in a block}
     &\rot[90]{05. Binary matrix rank test}
     &\rot[90]{06. Discrete Fourier transform}
     &\rot[90]{07. Non-overlapping template matching}
     &\rot[90]{08. Overlapping Template Matching}
     &\rot[90]{09. Maurer's universal statistics} 
     &\rot[90]{10. Linear complexity test} 
     &\rot[90]{11. Serial test}                       
     &\rot[90]{12. Approximate entropy test}          
     &\rot[90]{13. Cumulative sums test (forward) } 
     &\rot[90]{14. Cumulative sums test (reverse)}   
     &\rot[90]{15. Random excursions test}            
     &\rot[90]{16. Random excursions variant}
     &\phantom{~} 
     \\ \hline\\

{\tt CUE sampling, n=12}        && R & R & R & R & R & R & R & R & R & R & R & R & R & R & R & R \\ 
{\tt random bit sampling, n=53} && R & R & R & R & R & R & R & R & R & R & R & R & R & R & R & R \\
{\verb|n12_m14_s0_e0_pEFGH|}  && N & N & N & R & R & R & N & N & R & R & U & N & N & N & R & R \\ 
{\verb|n14_m14_s0_e0_pEFGH|}  && N & N & N & R & R & R & N & N & N & R & U & N & N & N & R & R \\
{\verb|n16_m14_s0_e6_pEFGH|}  && N & N & N & R & R & R & N & N & R & R & U & N & N & N & R & R \\ 
{\verb|n18_m14_s3_e6_pEFGH|}  && N & N & N & R & R & R & N & N & R & R & N & N & N & N & R & R \\
{\verb|n20_m14_s3_e0_pEFGH|}  && N & N & N & R & R & R & N & N & R & R & R & N & N & N & R & R \\
{\verb|n22_m14_s0_e0_pEFGH|}  && N & N & N & R & R & R & N & N & R & R & R & N & N & N & R & R \\
{\verb|n24_m14_s0_e0_pEFGH|}  && N & N & N & N & R & R & N & N & R & R & U & N & N & N & R & R \\
{\verb|n26_m14_s0_e0_pEFGH|}  && N & N & N & R & R & R & N & N & R & R & R & N & N & N & R & R \\
{\verb|n28_m14_s0_e0_pEFGH|}  && N & N & N & R & R & R & N & N & R & R & U & N & N & N & U & R \\
{\verb|n30_m14_s0_e0_pEFGH|}  && N & N & N & R & R & R & N & N & R & R & R & N & N & N & R & R \\
{\verb|n32_m14_s0_e0_pEFGH|}  && N & N & N & N & R & R & N & N & N & R & R & N & N & N & R & R \\
{\verb|n34_m14_s0_e0_pEFGH|}  && N & N & N & R & R & R & N & N & R & R & R & N & N & N & R & R \\
{\verb|n36_m14_s0_e0_pEFGH|}  && N & N & N & R & R & N & N & N & N & R & U & N & N & N & R & R \\
{\verb|n38_m14_s0_e0_pEFGH|}  && N & N & N & R & R & R & N & N & R & R & R & N & N & N & R & R \\
{\verb|n39_m14_s0_e0_pEFGH|}  && N & N & N & N & R & R & N & N & R & R & U & N & N & N & U & R \\
{\verb|n40_m14_s0_e0_pEFGH|}  && N & N & N & N & R & R & N & N & R & R & U & N & N & N & R & R \\
{\verb|n41_m14_s0_e0_pEFGH|}  && N & N & N & N & R & R & N & N & R & R & R & N & N & N & R & R \\
{\verb|n42_m14_s3_e6_pEFGH|}  && N & N & N & N & R & R & N & N & R & R & U & N & N & N & R & R \\
{\verb|n43_m14_s0_e0_pEFGH|}  && N & N & N & N & R & N & N & N & R & R & U & N & N & N & R & R \\
{\verb|n44_m14_s0_e0_pEFGH|}  && N & N & N & R & R & R & N & N & N & R & R & N & N & N & R & R \\
{\verb|n45_m14_s0_e0_pEFGH|}  && N & N & N & N & R & R & N & N & R & R & R & N & N & N & R & R \\
{\verb|n46_m14_s0_e0_pEFGH|}  && N & N & N & N & R & R & N & N & R & R & U & N & N & N & R & R \\
{\verb|n47_m14_s0_e0_pEFGH|}  && N & N & N & R & R & R & N & N & R & R & U & N & N & N & R & R \\
{\verb|n48_m14_s0_e0_pEFGH|}  && N & N & N & N & R & R & N & N & N & R & U & N & N & N & R & R \\
{\verb|n49_m14_s0_e6_pEFGH|}  && N & N & N & R & R & R & N & N & R & R & U & N & N & N & R & R \\
{\verb|n50_m14_s0_e0_pEFGH|}  && N & N & N & R & R & R & N & N & N & R & U & N & N & N & R & R \\
{\verb|n51_m14_s0_e0_pEFGH|}  && N & N & N & N & R & R & N & N & R & R & U & N & N & N & R & R \\
{\verb|n53_m14_s0_e0_pEFGH|}  && N & N & N & R & R & R & N & N & R & R & U & N & N & N & R & R \\[4pt]
{\verb|n53_m12_s1_e0_pABCDCDAB|} && N & N & N & R & R & R & N & N & R & R & U & N & N & N & R & R \\
{\verb|n53_m14_s2_e0_pABCDCDAB|} && N & N & N & N & R & R & N & N & R & R & U & N & N & N & U & U \\ 
{\verb|n53_m16_s0_e0_pABCDCDAB|} && N & N & N & N & R & R & N & N & R & R & U & N & N & N & R & R \\
{\verb|n53_m18_s0_e0_pABCDCDAB|} && N & N & N & N & R & R & N & N & R & R & N & N & N & N & U & R \\
{\verb|n53_m20_s0_e0_pABCDCDAB|} && N & N & N & N & R & R & N & N & N & R & U & N & N & N & R & R \\
{\verb|n53_m20_s1_e0_pABCDCDAB|} && N & N & N & R & R & R & N & N & R & R & U & N & N & N & R & R \\
\hline\hline
\end{tabular}
\caption{NIST random number tests on the Haar measure sampling for $n=12$, classical random bit sampling for $n=53$, 
random quantum circuit sampling with Qiskit for $n=12$ and  Google's random quantum circuit data. 
The letter `R' stands for random and `N' for non-random. 
         The letter `U' stands for some sub-tests that are random and the others non-random. The other results of
         the NIST random number tests on Google's random quantum circuits are found in the Supplementary Data.}
\label{Random_number_test}
\end{table*}
\end{widetext}
\end{document}